\documentclass[preprint,superscriptaddress,amsmath,amssymb,aps,physrev,prb,floatfix, fleqn]{revtex4-2} \usepackage{bm} 
\usepackage{hyperref}
\usepackage[english]{babel}
\usepackage[utf8]{inputenc}   
\usepackage{xspace}
\usepackage{tensor} 
\usepackage{array,color}
\usepackage{relsize,exscale}  
\usepackage{textcase}	
\usepackage{soul,color}		
\usepackage{calc} 
\usepackage{makeidx,showidx}
\usepackage{graphicx}
\usepackage{wrapfig}
\usepackage{textcomp}
\usepackage{siunitx}
\usepackage[version=3]{mhchem}
\usepackage{enumitem}

\newcommand{\erww} [1] {\ensuremath{\langle {#1} \rangle}}

\newcommand{\lsco} {{La$_{2-x}$Sr$_x$CuO$_4$}\@\xspace}

\newcommand{\ybco} {$\ce{YBa2Cu3O_{6+y}}$\@\xspace}

\newcommand{\ybcoE} {$\ce{YBa2Cu4O8}$\@\xspace}

\newcommand{\tc} {\ensuremath{T_{\mathrm c}}\@\xspace}

\newcommand{\tcmax} {\ensuremath{T_{\mathrm{c, max}}}\@\xspace}

\newcommand{\beq} {\begin{equation}}
\newcommand{\eeq} {\end{equation}}

\preprint{Bandur 2comp}
\begin{document}

\title{\textbf{Two-carrier description of cuprate superconductors from NMR}}
\author{Daniel Bandur}
\author{Abigail Lee}
\author{Jakob Nachtigal}
\author{Stefan Tsankov}
\author{Jürgen Haase}
\email{j.haase@physik.uni-leipzig.de}
\affiliation{University of Leipzig, Felix Bloch Institute for Solid State Physics, Linn\'estr.\@ 5, 04103 Leipzig, Germany}

\date{\today}

\begin{abstract}
{Cuprates currently hold the record for the highest temperature superconductivity at ambient pressure, but the microscopic understanding of these materials remains elusive. Here we utilize nuclear magnetic resonance (NMR) data of planar oxygen and copper from essentially all hole-doped cuprates to provide a universal phenomenology relating the NMR spin shifts, which measure the electronic spin polarization at a given nucleus, with the superconducting dome and maximum critical temperature. We demonstrate that there are two separate contributions to the spin shift at planar copper, only one of which is seen at oxygen, and associate them with two different carrier types. Upon disentangling these two components, their relative size is shown to determine not only the doping dependence of the superconducting dome, but also the variation in maximum superconducting critical temperature, $T_\mathrm{c}$, between different families. One of these components is independent of family and resides in the hybridized planar orbitals. The second component, in contrast, has a more three-dimensional character and encodes the differences between the families. It is thus related to the charge transfer gap and planar hole sharing. Our findings offer a key, universal insight which should prove useful in the continuing development of a comprehensive theory of the cuprates, as well as an indication of how it may be possible to engineer materials with higher critical temperatures.}
\end{abstract}


\maketitle

\section{Introduction}
\begin{figure}
\centering
\includegraphics[width=0.9\textwidth ]{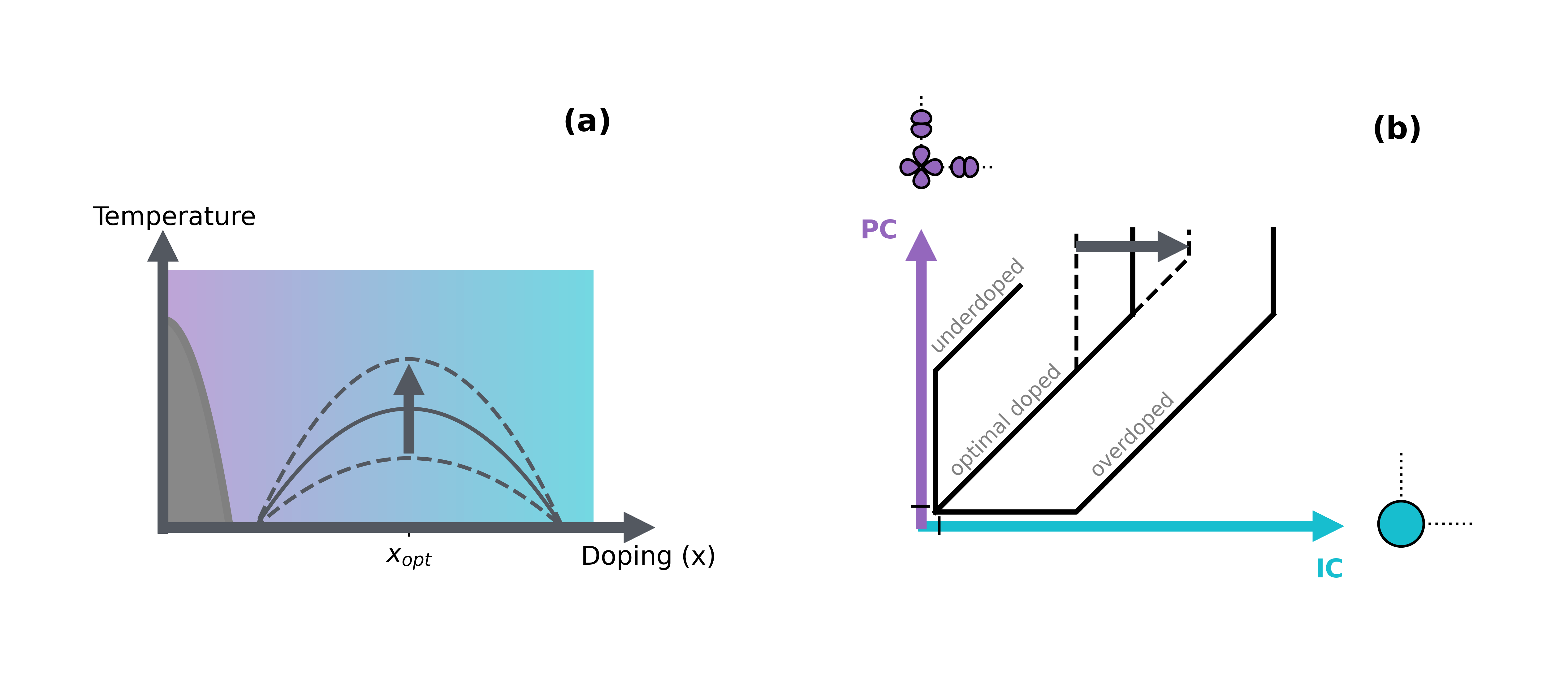}
\caption{(a) Phase diagram commonly used to discuss cuprate properties as a function of doping $x$. Antiferromagnetism is rapidly lost with increasing $x$ and conductivity sets in. Superconductivity appears with a maximum critical temperature ($T_\mathrm{c,max}$) atop the superconducting dome at optimal doping ($x_\mathrm{opt}$) that does not follow from $x$. \tcmax is however controlled by material dependent parameters not contained in the common phase diagram. The arrow indicates an additional parameter, with which  \tcmax increases.  
(b) Condensation behavior of the two components as inferred from their respective spin polarizations, sketched with temperature as an implicit parameter. Planar carriers (PCs, purple), involving hybridized Cu and O orbitals, are rather independent of family. Interplanar carriers (ICs, blue) show a clear family dependence. 
For underdoped samples, the spin polarization of both carriers begins to disappear at the pseudogap temperature. If the ICs are depleted, only the PCs continue. The opposite happens at high doping levels where the number of ICs can be larger. For optimal doping, both carriers appear matched, i.e. they condense together with a constant slope down to the lowest temperatures. A larger number of ICs at optimal doping favors a higher $T_\mathrm{c,max}$ as shown by the arrow. $T_\mathrm{c,max}$ appears ultimately limited by the universal PCs.}\label{fig:fig1}
\end{figure}

Since the discovery of cuprate high-temperature superconductivity \cite{Bednorz1986}, condensed matter physics has struggled with the understanding of these complex materials. Described in the common phase diagram, cf.\@ Fig.~\ref{fig:fig1}(a), as a function of doping the parent insulator evolves from its antiferromagnetic phase into a pseudogap and superconducting phase, and eventually into what is believed to be a metallic phase that does not superconduct. Unlike classical superconductors, for which a gap opens in a metallic density of states only at the critical temperature of superconductivity ($T_\mathrm c$) as the temperature is lowered, the still mysterious pseudogap phase shows lost states already far above $T_\mathrm c$. 
The maximum critical temperature, $T_\mathrm{c,max}$, observed atop the superconducting dome, is of great interest in the field, but does not follow from the simple phase diagram. Rather, it depends on the material family \cite{Uemura1989}, and 
it was noted that the apical oxygen distance to the planar copper correlates with $T_\mathrm{c,max}$ \cite{Ohta1991,Pavarini2001,Mahony2022}, which can involve the rather large Cu $4s$ orbital. It was also shown that the same family dependence is reflected in the sharing of the inherent hole between planar Cu ($n_\mathrm{d}$) and O ($n_\mathrm{p}$), which can be determined from the quadrupole splitting measured with nuclear magnetic resonance (NMR) at the planar nuclei \cite{Zheng1995h,Haase2004}. In fact, the empirical relation, $T_\mathrm{c,max} \approx 200 \text{K} \cdot 2 n_\mathrm p$, was found for the hole-doped cuprates \cite{Rybicki2016}. $T_\mathrm{c,max}$ is nearly proportional to the planar O $2p_\sigma$ hole content at optimal doping (while $n_\mathrm d+2n_\mathrm p \approx 1+x$, as expected from chemistry). Recent progress with advanced computations (DMFT) endorse these findings  \cite{Kowalski2021,Bacq2025}. Furthermore, the relation was also shown to explain the old conundrum of how pressure can increase $T_\mathrm{c, max}$ beyond what can be achieved by chemical doping: in such a case pressure also changes the sharing of the hole between planar Cu and O accordingly \cite{Jurkutat2023}. One then wonders whether this intriguingly simple dependence of $T_\mathrm{c,max}$ on $n_\mathrm p$ and $n_\mathrm{d}$, and the family dependence in general, has its counterpart in the magnetic NMR data, which influenced the field profoundly early on \cite{Slichter2007}.

Here we uncover a new universal scaling between the magnetic \emph{axial} shift of planar Cu and the magnetic shift of planar O, and an additional magnetic shift on planar Cu that is not present at planar O. The magnetic shift contribution common to both nuclei is assigned to planar carriers, given the involved hyperfine couplings, while a second spin component, likely from interplanar carriers, determines the family dependent remainder of the Cu shift. Importantly, both spin components apparently act together to set $T_\mathrm{c}$ and the superconducting dome, as well as $T_\mathrm{c,max}$, which is sketched in Fig.~\ref{fig:fig1}(b). While the focus here is on the shifts, nuclear relaxation seems to follow from the same scenario, although the detailed understanding requires theory.

This simple new scenario across all hole-doped cuprate families shows that the understanding of cuprate superconductivity needs a broader view. A sole focus on the generic phase diagram or certain cuprate families is insufficient, as we already know from the importance of the planar charge sharing between Cu and O.

\section{NMR Shifts}
{\flushleft We begin with a short review of the cuprate shifts, specifically of the CuO$_2$ plane.}

A material in a magnetic field, $B_0$, responds with an electronic spin polarization, $\erww{S_z}$, given by the uniform electronic spin susceptibility, $-\gamma_e \hbar \erww{S_z}= \chi B_0$. In the case of isotropic spin, this causes a relative NMR frequency shift, the spin shift, 
\begin{equation}\label{eq:1}
K_\alpha(T) = A_\alpha \cdot \chi(T),
\end{equation}
where $A_\alpha$ is the anisotropic hyperfine constant for that particular ion. In ordinary metals, $\chi$ is proportional to the density of electronic states (DOS) near the Fermi surface (selected by the Fermi function at temperature $T$). This is the positive and temperature-independent Pauli susceptibility or a metal's Knight shift. If such a material becomes superconducting at a critical temperature $T_\mathrm c$, the shift begins to decrease  and vanishes towards low temperatures for spin singlet pairing \cite{Yosida1958,Schrieffer2018}. It is important to note that $\chi(T)$ in \eqref{eq:1} leads to proportional changes of the shifts at all nuclei, only their sizes due to the respective hyperfine constants differ. For more complicated hyperfine scenarios, e.g.\@ even in simple metals with more than one band, this proportionality can be lost if more than one spin component is present \cite{Carter1976}.
\begin{figure}
\centering
\includegraphics[width=0.575\textwidth ]{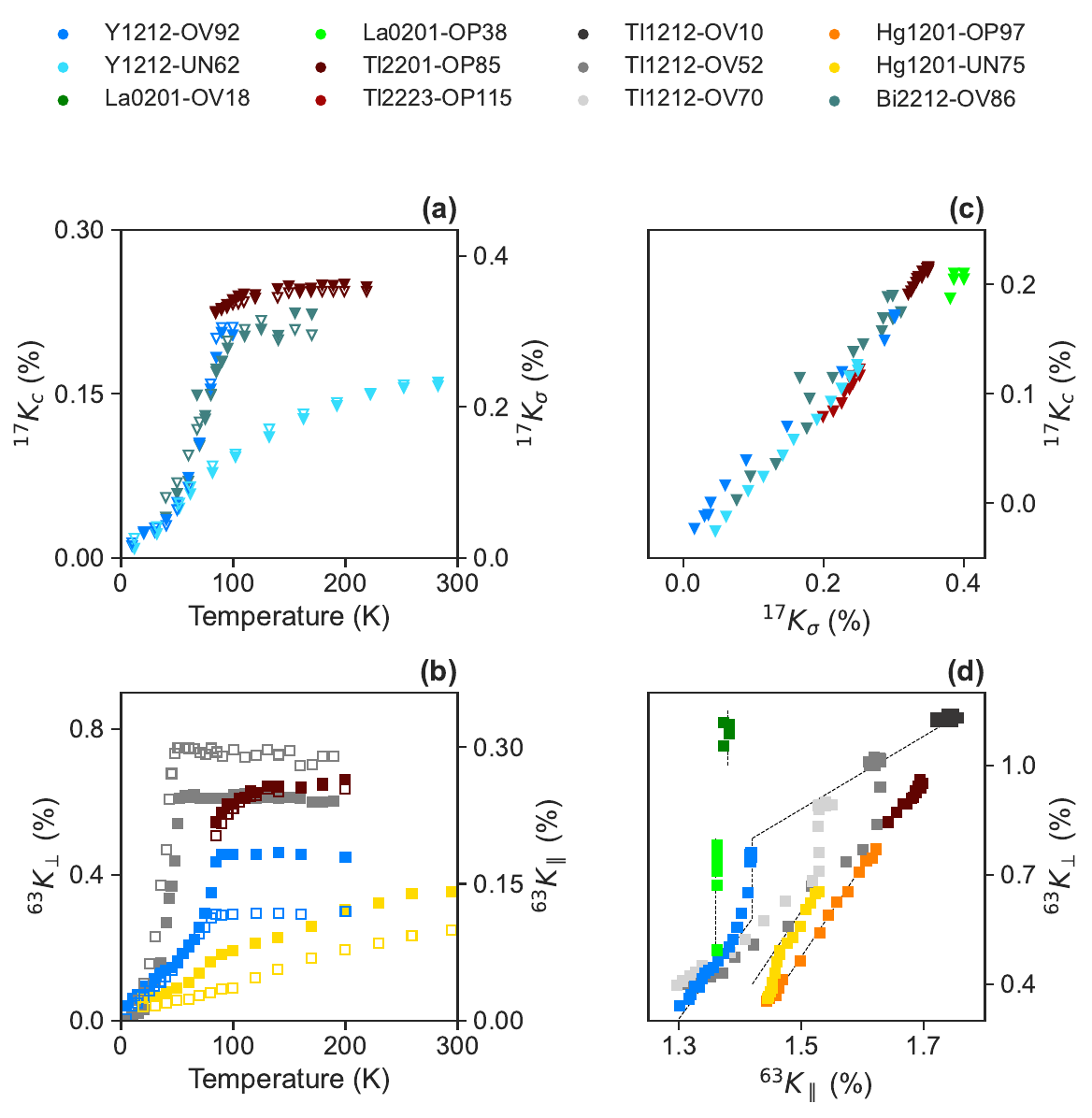}
\caption{Typical examples of temperature-dependent planar shifts. (a) Oxygen  for two directions of the external field: perpendicular to the $p_\sigma$ bond along the crystal $c$-direction (${^{17}K}_c$, full symbols), and along the $p_\sigma$ bond (${^{17}K}_\sigma$, open symbols). The shifts are nearly proportional to each other as can be seen in (c). The situation is very different for planar Cu (b). Here the temperature-dependence of the in-plane shift (${^{63}K}_\perp$, full symbols) and that along the crystal $c$-axis (${^{63}K}_\parallel$, open symbols) are rather different. However, one observes nearly straight line segments with 3 different slopes (d) as indicated by the thin lines  \cite{Haase2017}. Low temperature shift offsets have been subtracted in (a), (b).}\label{fig:fig2}
\end{figure}
\subsection{Brief Review of Planar Cuprate Shifts}
Early tests of eq.\@ \eqref{eq:1} for the cuprates were performed on two materials, YBa$_2$Cu$_3$O$_{6.63}$ and \ybcoE  \cite{Takigawa1991,Bankay1994}. Since the temperature dependence of the planar Cu shift, $^{63}K_\perp(T)$ (magnetic field lies in the CuO$_2$ plane) is rather similar to that of planar O, $^{17}K_{\alpha}(T)$ (all field directions), a single component view appeared appropriate. However, the hyperfine scenario for Cu had to be more complex since the shift with the field along the crystal $c$-axis was found to be rather temperature-independent, $\Delta_\mathrm{T} {^{63}K}_\parallel \approx 0$, above and below $T_\mathrm{c}$ for these materials. In a single band view, one then concluded that $^{63}K_\alpha = (A_\alpha + 4B) \chi$ with $A_\parallel \approx -4 B$, and $^{17}K_\alpha = 2C_\alpha \chi$ (O is situated between two Cu atoms, each Cu atom is surrounded by 4 Cu neighbors) \cite{Slichter2007}. When some other materials showed a temperature-dependent ${^{63}K}_\parallel$, it was ascribed to variations in the hyperfine constant $B$ \cite{Kitaoka1991}. Additionally, uncertainties remained due to the often large shift distributions, in particular at low temperatures, as well as the lack of the precise knowledge of the Meissner fraction \cite{Barrett1990b}.

Later was it shown that La$_{1.85}$Sr$_{0.15}$CuO$_{4}$ cannot be understood in such simple terms \cite{Haase2009b}. More evidence came from experiments on other systems over the years \cite{Haase2012,Rybicki2015}. Finally, a thorough analysis of all planar Cu data available in the literature \cite{Haase2017} showed that even shifts observed at the same Cu nucleus, but different directions of the field, $\Delta_\mathrm{T}{^{63}K}_\parallel$ and $\Delta_\mathrm{T}{^{63}K}_\perp$, are in general not proportional to each other. However, one observes only 3 different ratios $\Delta_\mathrm{T}{^{63}K}_\parallel/\Delta_\mathrm{T}{^{63}K}_\perp$ in given temperature intervals. This behavior cannot be understood with a single spin component, but also demands a rather universal Cu hyperfine scenario, in contrast to assumptions mentioned above. Examples of shifts are shown in Fig.~\ref{fig:fig2}. 

Surprisingly, the planar $^{17}$O shifts (and relaxation) are rather simple, as a recent literature overview showed \cite{Nachtigal2020}. They are essentially independent of family and their higher temperature behavior is explained with a temperature-independent but doping dependent pseudogap at the Fermi level in an otherwise universal electronic density of states. In other words, the planar O data can be understood with just two parameters: a pseudogap (with a size set by doping only) and a single electronic spin component. 

In a recent summary, this apparent shift dichotomy between Cu and O across all materials is discussed \cite{Avramovska2022}.

\subsection{New universal scaling law}
\begin{figure}
\centering
\includegraphics[width=0.5\textwidth ]{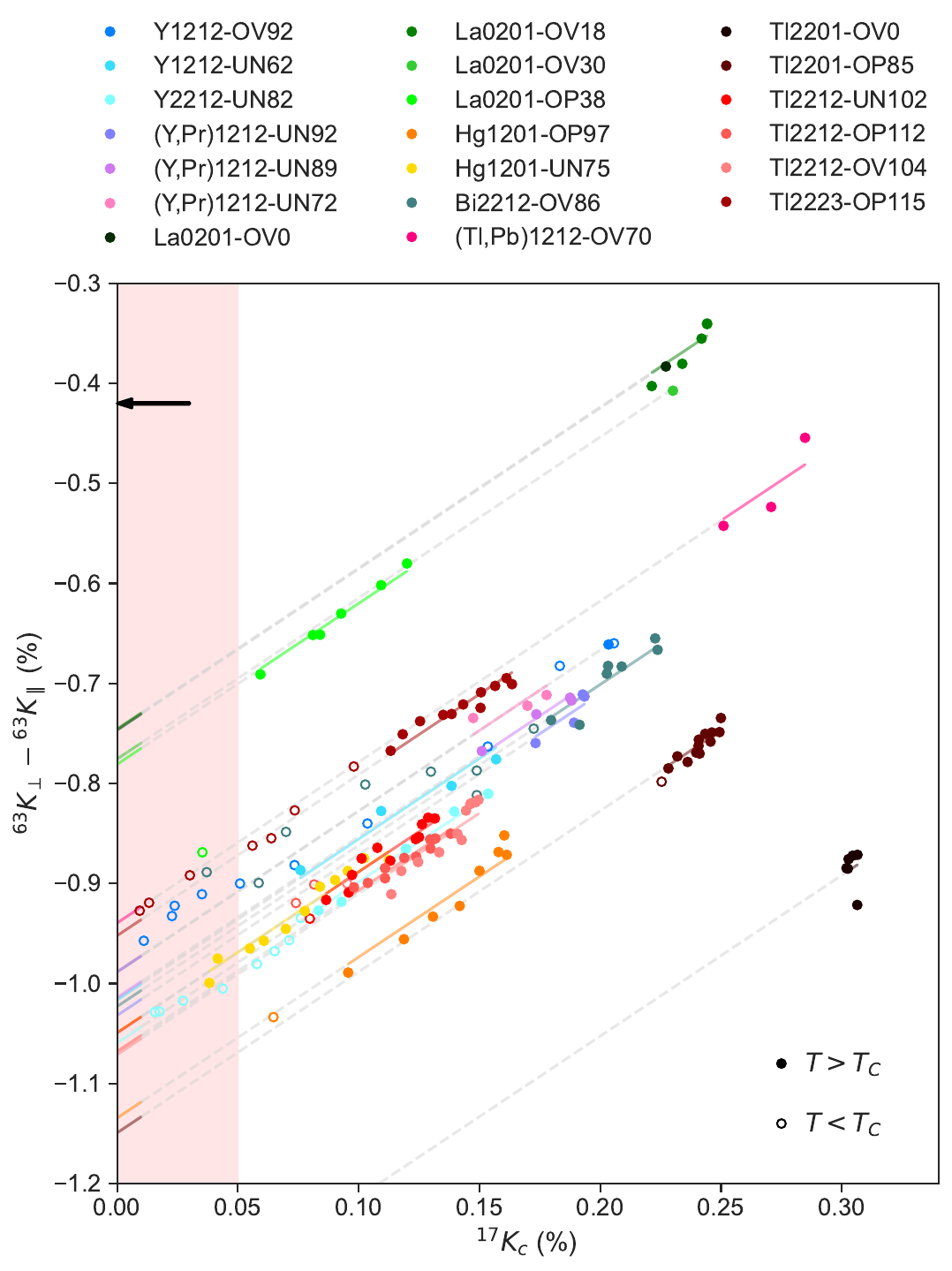}
\caption{Total planar Cu axial shift plotted against the shift of planar O. The small orbital shift for O was removed \cite{Avramovska2022b}, but for planar Cu the full shifts were used. For a given doping and family, the Cu axial shift is proportional to the planar O shift. More data in the red shaded, low-temperature region would be desirable. Note that the relation holds above and below $T_\mathrm{c}$, full and open circles respectively. Full lines are fits to the data, dashed lines their extension to lower temperatures. The low-temperature vertical drop for La$_{1.85}$Sr$_{0.15}$CuO$_4$ is due to a change in the axial shift and not dictated by $\delta$  ($\delta$ is a temperature-independent offset, see main text). The arrow denotes the assumed orbital shift anisotropy ${^{63}\Lambda}= -0.42\%$ \cite{Renold2003}.}\label{fig:fig3}
\end{figure}

While searching for a simple phenomenology in the cuprate shifts, we uncovered an unexpectedly robust property that reveals itself in Fig.~\ref{fig:fig3}, and is described by,
\begin{equation}\label{eq:scaling}
{^{63}K}_\mathrm{ax}(T) \equiv {^{63}K}_\perp(T) - {^{63}K}_\parallel(T) \approx 1.6 \cdot {^{17}K}_\mathrm{c}(T) + \delta.
\end{equation}
This means that the Cu axial shift, i.e.\@ the difference of the shifts with magnetic field in the plane and perpendicular to it (the shift tensor is in-plane symmetric), has the same temperature dependence as that of planar O (for any direction of the field, cf.\@ Fig.~\ref{fig:fig2}). The universal proportionality constant of about 1.6 applies to ${^{17}K}_\mathrm{c}$, the most often measured direction. La$_{1.85}$Sr$_{0.15}$CuO$_4$ is special in that the sudden drop at lower temperatures in Fig.~\ref{fig:fig3} is not due to $\delta$, rather due to ${^{63}K}_\perp$, so the large high-temperature offset is not just given by $\delta$. This was studied before \cite{Haase2009b}, and is likely true for the other \lsco materials. This brings $\delta$ closer to the values for the other cuprates. Relation \eqref{eq:scaling} resolves the mystery of the missing scaling of the bare shifts, as there is obviously another, family and temperature-dependent shift of planar Cu that is not present on planar O. For \lsco this component is axial while for all other cuprates it is isotropic.

Note that the temperature dependence of a shift in the pseudogap phase is rather sensitive to doping, i.e., data from different doping levels would not lead to straight lines in Fig.~\ref{fig:fig3}. This shows that all literature data, even from different sources and preparation processes, are rather reliable in that sense, and varying linewidths do not seem to matter. Relation \eqref{eq:scaling} holds whether the shifts become temperature-dependent at \tc (overdoped materials) or already at much larger temperatures in the presence of the pseudogap. One concludes that this is the action of a single temperature-dependent electronic spin component, while the remainder of the Cu shift is attributed to a second electronic spin component not seen by the O nucleus.

For \lsco, and \ybco above about 80K, we note that $^{63}K_\parallel$ is largely temperature-independent ($\Delta_T{^{63}K}_{\parallel}\approx 0$), i.e.\@ the temperature dependences of ${^{63}K}_\mathrm{ax}$ and ${^{63}K}_\perp$ are the same (these systems defined the old picture).
Below \SI{80}{K}, ${^{63}K}_\parallel$ of \ybco suddenly becomes temperature-dependent, but shows no change in slope in Fig.~\ref{fig:fig3} due to the isotropic origin.

The temperature-independent offset $\delta$ is dominated by ${^{63}K}_\parallel$ for most systems, as one can see already with Fig.~\ref{fig:fig2}, but it also has small contributions from ${^{63}K}_\perp$ (minor variations also originate in the planar O shifts). One would expect it to be given by the orbital shift contributions. This will be addressed further below.

To conclude, it is the planar Cu axial shift and the planar O shift that show single-component behavior across all cuprates. It is predominantly planar Cu that is affected by another temperature-dependent spin component. This is an overarching, very robust scenario based on all available data in the literature.
\section{Two-component scenario}
Since it is not clear how one should describe the shifts in this more complicated situation, we establish a simple phenomenology that carries the essential relations in terms of two independent carriers. 
We ascribe the straight lines in Fig.~\ref{fig:fig3} to an electronic spin component given by $\chi_\mathrm{PC}(T)$. It sets the Cu axial shift, ${^{63}K}_\mathrm{ax}(T)={^{63}K}_\perp(T)-{^{63}K}_\parallel(T)$, and is the sole term for the planar O shift, e.g.\@ ${^{17}K}_\mathrm{c}$(T). We introduce a new hyperfine coefficient, $R_\alpha$, for planar Cu. For planar O we keep the old coefficient, $2C_\alpha$. We then have,
\begin{align}
{^{63}K}_\mathrm{ax}(T) &= ({R}_\perp-{R}_\parallel) \cdot\chi_\mathrm{PC}(T)\label{eq:axshift}\\
{^{17}K}_\mathrm{c}(T)&= 2C_\mathrm{c}\cdot\chi_\mathrm{PC}(T)\label{eq:oxshift}.
\end{align}
In view of Fig.~\ref{fig:fig3} we write,
\begin{equation}\label{eq:scaling2}
{^{63}K}_\mathrm{ax}(T) = \frac{R_\perp-{R}_\parallel}{2C_\mathrm{c}} \;{^{17}K}_\mathrm{c}(T) + \delta,
\end{equation}
where $R_\perp, R_\parallel$, and $C_\mathrm{c}$ are the corresponding effective, anisotropic hyperfine coefficients, and $\delta$ is the temperature-independent offset (note that for \lsco this is only the low temperature offset). Then,
\begin{equation}
\frac{R_\perp-R_\parallel}{2C_\mathrm{c}} \approx 1.6 \;(\pm 0.2).
\end{equation}
Clearly, since $\chi_\mathrm{PC}$ drives the planar Cu axial and the planar O temperature-dependent shifts, independent of family, it will be due to spin from expected planar carriers (PC). In fact, based on the known contributions from the hyperfine coefficients \cite{Huesser2000,Slichter2007}, $\chi_\mathrm{PC}(T)$ is readily assigned to the intrinsic hole in the $3d(x^2-y^2)$ orbital that is hybridized with planar O. One is inclined to use the established hyperfine coefficients for it, i.e. for planar Cu, \@ $R_\alpha = (A_\alpha+4B)$, with $-A_\parallel \approx 4B$, to make sure ${^{63}K}_\parallel$ is temperature-independent for \lsco and \ybco (at higher temperatures). The additional shift component on planar Cu we attribute to a second susceptibility, $\chi_\mathrm{IC}$, and write,
\begin{equation}\label{eq:twoSpins}
{^{63}K}_\alpha(T) = R_{\alpha} \chi_\mathrm{PC}(T) + S_\alpha \chi_\mathrm{IC}(T).
\end{equation}
The hyperfine coefficient $S_\alpha$ is isotropic for most cuprates and axial for \lsco. $S_\alpha$ was not uncovered with first principle cluster calculations that focused only on the planar structure \cite{Huesser2000}, but band-structure calculations point to such contributions \cite{Pavarini2001}. The label IC reminds us of the interplanar nature of this spin component.\par\medskip

While the set of data in Fig.~\ref{fig:fig3} is convincingly large, there are much more planar Cu data available for which ${^{17}K_c}$ was not measured due to the necessary isotope exchange. We focus now on what one can learn from these additional planar Cu shifts. A collection of all literature $^{63}$Cu shifts was published recently, and a convenient way of looking at the many sets of data (from 19 families) is a plot of ${^{63}K}_\perp(T)$ vs.\@ ${^{63}K}_\parallel(T)$ \cite{Haase2017}. A few examples are included in Fig.~\ref{fig:fig2}(b, d). Most interesting is the fact that such a plot consists of a set of  straight line segments. This means that, as a function of temperature or doping, changes in both shifts, $\Delta {^{63}K}_\perp$ and $\Delta {^{63}K}_\parallel$, are related to each other, with only 3 different slopes, 
\begin{equation}\label{eq:slopes}
\Delta {^{63}K}_\perp/\Delta {^{63}K}_\parallel = \sigma_i, \;\;\; \sigma_{1,2,3}\approx{\infty, \frac{5}{2}, 1}.
\end{equation}
Thus, at certain points in temperature (or doping), one finds rather sudden changes between those slopes. It should be stressed that the different slopes cannot be due to sudden changes in the hyperfine coefficients since we do not observe any discontinuities in the shifts themselves. Additionally, the spin susceptibility is isotropic, i.e. it is generally independent on the direction of the magnetic field (as seen from the scaling of the planar oxygen shifts in all field directions). Comparing the 3 slopes with the two components in \eqref{eq:twoSpins}, one concludes,
\begin{align}
\sigma_1& \rightarrow \Delta \chi_\mathrm{IC} \approx 0\label{eq:s1}\\
\sigma_2& \rightarrow \Delta \chi_\mathrm{IC} \approx \frac{8B}{3S} \Delta \chi_\mathrm{PC}\label{eq:s2}\\
\sigma_3&  \rightarrow \Delta\chi_\mathrm{PC} \approx 0.\label{eq:s3}
\end{align}
Thus, the 3 slopes in \eqref{eq:slopes} are simply the result of the independent behavior of the two spin components. If only one of the components is temperature-dependent we find slope $\sigma_1$ or $\sigma_3$. If both change together it occurs with the unique ratio $\sigma_2$. Note that switching between these slopes can occur near $T_\mathrm{c}$ (overdoped samples), as well as temperatures comparable to the size of the pseudogap. Furthermore, such changes occur at presently unforeseen temperatures. The slope $\sigma_2$ \eqref{eq:s2} might suggest that $S= 8B/3$, but that is not necessarily the case due to the unknown relative size of the susceptibilities.

\begin{figure}[t]
\centering
\includegraphics[width=0.6\textwidth ]{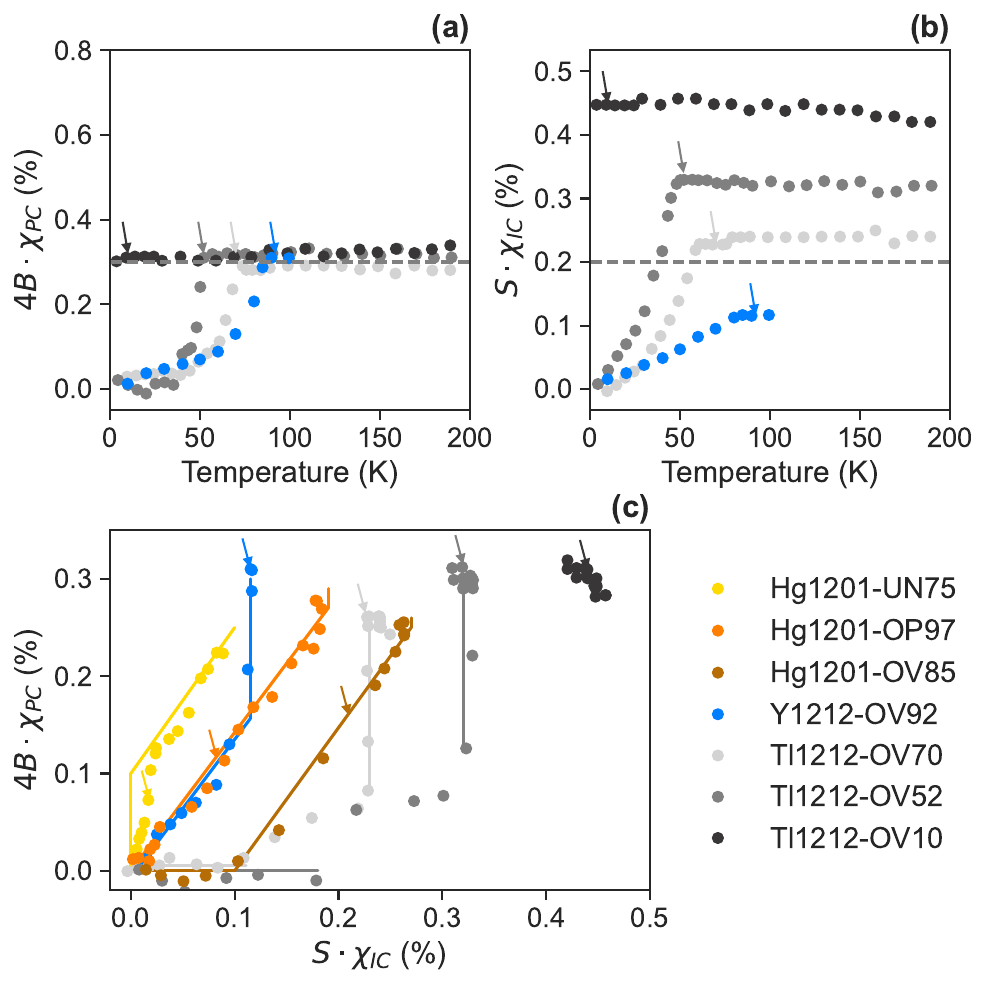}
\caption{Examples of extracted spin components for a few slightly underdoped and overdoped cuprates (arrows denote bulk $T_\mathrm{c}$). Low temperature spin shifts have been subtracted. (a) $\chi_\mathrm{PC}$ has the same high temperature value; the dashed line indicates this metallic shift. Note that $\chi_\mathrm{PC}$ is rather material and doping independent (at higher temperatures), and we observe a rather sharp drop at $T_\mathrm{c}$ (overdoped). (b) $\chi_\mathrm{IC}$ shows a different temperature dependence; the dashed line indicates the metallic shift for an isotropic DOS matched to that of the PC according to \eqref{eq:s2}. $\chi_\mathrm{IC}$ grows with doping; its condensation can occur at lower temperatures and the drop is less rapid. (c) Both spin components from (a) and (b) plotted against each other. At condensation either only PCs condense (vertical slope), or PC and IC condense together (slanted slope). If one of the carriers is fully condensed, the other can continue, leading to sudden changes of slopes: vertical (lower doping) or horizontal (higher doping). At optimal doping both carriers disappear together at a fixed rate.}\label{fig:fig4}
\end{figure}

For overdoped materials (see e.g. Hg1201-OV85, Tl1212-OV70, Fig. \ref{fig:fig4}), the shifts typically consist of two temperature-independent (metal-like) components. At the bulk \tc, condensation begins with $\chi_\mathrm{PC}$, while $\chi_\mathrm{IC}$ can follow immediately or join at lower temperatures. There may be two different condensation temperatures, with $T_{\mathrm c}^\mathrm{IC} \leq T_{\mathrm c}^\mathrm{PC}$. Upon lowering the temperature further, another change in slope to $\sigma_3$ indicates that $\Delta\chi_\mathrm{PC} \approx 0$, while $\chi_\mathrm{IC}$ continues to decrease. Again, this reminds one of different condensation behaviors (similar to different pairing scenarios).

Interestingly, optimal doping seems to demand that $\sigma_2$ \eqref{eq:s2} holds down to the lowest temperature (see e.g. Hg1201-OP97, Y1212-OV92, Fig. \ref{fig:fig4}). Once $\sigma_2$ is assumed, there will be no switch to $\sigma_1$ or $\sigma_3$ at lower temperatures, i.e. $\chi_\mathrm{IC}$ shows the same condensation behavior as $\chi_\mathrm{PC}$. Note that even for optimal doping it is always $\chi_\mathrm{PC}$ that first becomes temperature-dependent, and $\chi_\mathrm{IC}$ follows only at lower temperatures.

For underdoped samples (see e.g. Hg1201-UN75, Fig. \ref{fig:fig4}), it is not clear from the data how exactly $\chi_\mathrm{IC}$ is affected by the pseudogap. However as long as $\sigma_2$ \eqref{eq:s2} holds, both components have the same temperature dependence. In underdoped samples, this slope is typically observed up to the highest temperatures. At lower temperatures there is a change to slope $\sigma_1$ ($\Delta\chi_\mathrm{IC} \approx 0$), which is the opposite behavior compared to the overdoped samples.

It is clear from Fig.~\ref{fig:fig4}(b) that $\chi_\mathrm{IC}$ grows with doping, even beyond the closing of the pseudogap. $\chi_\mathrm{PC}$, on the other hand, stays, apart of the effect of the pseudogap, relatively fixed, independent of doping or family (see Fig.~\ref{fig:fig4}(a)). Due to the doping dependence of $\chi_\mathrm{IC}$, the measured spin shift in underdoped samples is typically dominated by $\chi_\mathrm{PC}$, while for overdoped samples it is $\chi_\mathrm{IC}$. Additionally, there is a family dependence in $\chi_\mathrm{IC}$. At optimal doping it appears that some families have a much larger $\chi_\mathrm{IC}$ than others. All families with a high \tcmax also have a large $\chi_\mathrm{IC}$, while those families with lower \tcmax have smaller $\chi_\mathrm{IC}$. 

With this information we can inspect \lsco again. This system lacks an isotropic spin shift on copper, but behaves similarly in terms of the planar carriers. In fact, the large positive offset in Fig.~\ref{fig:fig3} begins to disappear near \tc, while the low-temperature shift is similar to that of all the other systems (${^{63}K}_\perp (T \rightarrow 0) \approx 0.35\%$). From the overall phenomenology it then appears that \lsco has in addition to $\chi_\mathrm{PC}$, also $\chi_\mathrm{IC}$ with constant DOS above about \tc, however with predominantly axial symmetry. It also does not affect planar O significantly and begins to disappear only at lower temperature. This is behind the observation in \lsco that the planar Cu shift drops while the shift of planar O has already vanished below \tc \cite{Haase2009b}.


\paragraph*{Note on orbital shift contributions: }
The planar O orbital shift is small and in agreement with first principle calculations \cite{Renold2003}. Uncertainties do not interfere with the spin shift analysis. This is different for planar Cu with a hole in the $3d(x^2-y^2)$ orbital \cite{Pennington1989}. Due to the hybridization with planar O, the isolated ion orbital shift anisotropy is greatly reduced from a factor of approximately 4 to perhaps 2.4 \cite{Renold2003}, a value that is rather reliable.

Experimentally, ${^{63}K}_\perp(T)$ reaches a nearly common low-temperature value of about 0.35 to 0.40\%, and first principle calculations reported later that ${^{63}K}_\mathrm{L\perp} \approx 0.30\%$ \cite{Renold2003}. For ${^{63}K}_\parallel$ the situation is very different. There is no common low-temperature shift for the cuprates, and the same first principles calculations predict ${^{63}K}_\mathrm{L\parallel} = 0.72\%$ \cite{Renold2003}, in stark disagreement with even the \lsco temperature-independent shift of $\sim 1.3\%$. Uncertainties from large linewidths, the Meissner fraction, and a finite carrier lifetime may explain the small differences for ${^{63}K}_\mathrm{L\perp}$, but not for ${^{63}K}_\mathrm{L\parallel}$.

Since an orbital shift pair of ${^{63}K}_\mathrm{L\perp}=0.30\%$ and ${^{63}K}_\mathrm{L\parallel}=0.72\%$ is reasonable by all accounts, we indicated this value with an arrow at the ordinate in Fig.~\ref{fig:fig3}, i.e., we define the difference as ${^{63}\Lambda}\equiv{^{63}K}_\mathrm{L\perp}-{^{63}K}_\mathrm{L\parallel} = -0.42\%$.
The temperature-independent offset $\delta$ in Fig.~\ref{fig:fig3} is defined by \eqref{eq:scaling2} and thus contains this orbital shift. 
$\delta$ mainly changes between families; there is perhaps only a small doping dependence (see e.g. Tl2201, Fig \ref{fig:fig3}). 

This temperature-independent shift $\delta$ with its large anisotropy defies its identification as orbital shift (electron doped materials that are very similar in chemistry are much closer to the first principle calculations). Furthermore, the presence of new $s$-like states should not contribute to orbital moments, although $\delta$, in particular with the axis parallel to the field, has a similar family dependence as the spin shift $\chi_\mathrm{IC}$. 

\section{Discussion and Conclusion}
The newly discovered, very robust scaling between the planar Cu \emph{axial} shift and any planar O shift in essentially all cuprates resolves the long-standing NMR shift conundrum, i.e.\@ the missing scaling between the bare Cu and O (and other) shifts. It shows that the family independent planar O shift that is generic to the cuprates is present in full extent (with respect to the hyperfine coefficients) as Cu axial shift. In addition, most cuprates are found to have an additional planar Cu isotropic shift, except for \lsco where the second contribution is also axial. 

This scaling can be interpreted by two temperature-dependent spin susceptibilities, $\chi_\mathrm{PC}(T)$ and $\chi_\mathrm{IC}(T)$. It appears that $\chi_\mathrm{PC}(T)$ is a family-independent property of the CuO$_2$ plane, i.e.\@ it only depends on doping, with the corresponding spin polarization arising from hybridized Cu $3d(x^2-y^2)$ and O $2p_\sigma$ orbitals. At higher doping levels, $\chi_\mathrm{PC}(T)$ becomes temperature and doping independent (metal-like) and changes only at $T_\mathrm{c}$. For lower doping levels, i.e.\@ in the presence of the pseudogap, $\chi_\mathrm{PC}$ already decreases at a higher temperature. Since this behavior is similar for all cuprates, $\chi_\mathrm{PC}$ is not affected by the family dependent sharing of charge between Cu and O in the plane that sets $T_\mathrm{c,max}$ \cite{Jurkutat2014,Rybicki2016}. 
The second electronic spin from $\chi_\mathrm{IC}(T)$, on the other hand, is strongly family dependent.
It sets the additional shift at planar Cu and is isotropic for most materials (except \lsco). Unlike $\chi_\mathrm{PC}$, which saturates to a similar high-temperature value, independent of doping, $\chi_\mathrm{IC}$ grows with doping even after the pseudogap is closed. $\chi_\mathrm{IC}$ is also metal-like and affected by a pseudogap (it is not certain from the data whether it is the same pseudogap as that seen by $\chi_\mathrm{PC}$). Note that a doping dependent density of states at the Fermi surface as for $\chi_\mathrm{IC}$ is a signatue of 3 dimensional metals. A doping independent density of states at the Fermi surface, observed for $\chi_\mathrm{PC}$, is expected for 2 dimensional metals.

Both susceptibilities can drop (from carrier condensation) rapidly at certain temperatures, apparently independently. Above optimal doping, condensation begins with $\chi_\mathrm{PC}$ at \tc and $\chi_\mathrm{IC}$ joins the condensation at slightly lower temperatures, up to about 15 degrees. This may explain the not so sharp transition into the superconducting state often ascribed to inhomogeneity. Below \tc, it appears that $\chi_\mathrm{PC}$ drops more rapidly than $\chi_\mathrm{IC}$, perhaps pointing to different gap symmetries \cite{Mueller1995}. 

Interestingly, the superconducting dome relates to the match or mismatch of the two susceptibilities. At optimal doping both fall together to the lowest temperature with a fixed rate, while for under- or overdoped samples one of the two susceptibilities is depleted earlier. The family dependent highest transition temperature, $T_\mathrm{c,max}$, which occurs at optimal doping, seems to be related to the two components. It appears as if optimally doped samples with a high amount of $\chi_\mathrm{IC}$ also have a high $T_\mathrm{c,max}$. As such, $\chi_\mathrm{IC}$ is also reminiscent of the correlation between \tcmax and the apical oxygen distance, which relates to the involvement of the copper 4s orbital \cite{Pavarini2001}. For a high $T_\mathrm c$, it is also favorable if $\chi_\mathrm{IC}$ is matched to $\chi_\mathrm{PC}$ according to \eqref{eq:s2}. If this phenomenology holds, a larger planar $\chi_\mathrm{PC}$ might be necessary for cuprates with much higher $T_\mathrm{c}$. In other words, $T_\mathrm{c,max}$ of the cuprates seems ultimately set by the CuO$_2$ plane, which may explain why the cuprate $T_\mathrm{c,max}$ is somehow limited currently. 

It was noted before \cite{Nachtigal2020,Tallon2022} that $\chi_\mathrm{PC}$ (at planar O) explains Loram's specific heat in terms of the temperature dependence \cite{Loram1998, Tallon2022}. Given that $\chi_\mathrm{IC}$ is rather close in its temperature dependence, in particular in the pseudogap regime, where both spin components decrease together, they probably cannot be separated in the specific heat. The increase in $\chi_\mathrm{IC}$ on the overdoped side may be behind the increase in DOS reported with specific heat data \cite{Loram1998, Michon2019}. Nuclear relaxation of planar O is in agreement with the specific heat \cite{Nachtigal2020}, and both components seem to be involved in setting the Cu relaxation, as the anisotropy decreases with doping as expected from to the growth of $\chi_\mathrm{IC}$. In addition, \lsco has a significantly larger relaxation when the field is in the CuO$_2$ plane, likely due to the additional axial term.

The thermal excitations leading to $\chi_\mathrm{PC,IC}$ appear metallic \cite{Barisic2019} even in the underdoped region of the phase diagram. However, even at rather high doping levels ($x \sim 0.3$) the response is still not what is expected from simple classical metals \cite{Bandur2025, Li2023}. Stripe-like correlations seem to be ubiquitous for $\chi_\mathrm{PC}$ \cite{Bandur2025,Tranquada1995}, which points to the role of lattice degrees of freedom in the susceptibility \cite{Keller2008,Bianconi1996}. The isotope effect for planar O might not be surprising in this scenario \cite{Bussmann2024}. It may also be possible that a phase separation between two components is the reason for the ubiquitous inhomogeneity observed in the cuprates \cite{Gorkov2001}.

The two carriers remind one of a two-band scenario \cite{Suhl1959, Bussmann2024, Matt2018}, which would readily explain differences in the critical temperature $T_\mathrm{c}$ between the two components as well as the importance of the doping- and family dependent density of states of the carriers for superconductivity. Two condensation temperatures can be observed in absence of a strong inter-band scattering \cite{Suhl1959}. Two carriers may also be required in loop current scenarios \cite{Weber2014, Bourges2022}.

With the identification of $\chi_\mathrm{PC,IC}$, the temperature-independent shift offset $\delta$ becomes clearer. It varies between the different samples and is not in agreement with simple orbital shift scenarios \cite{Renold2003,Zheng1995h}. This offset is found to be correlated with $\chi_\mathrm{IC}$ and to the oxygen hole content $n_\mathrm p$ (and thus \tcmax \cite{Rybicki2016}), which makes an identification as orbital shift questionable. If $\delta$ were caused by spin it may require a negative spin polarization \cite{Haase2017}.\par\medskip

To conclude, we described a magnetic shift scenario of planar Cu and O that carries a universal doping dependent and family independent component, as well as a second, family and doping dependent component. The interplay of the condensation of the two associated carriers describes the superconducting dome, i.e.\@ $T_\mathrm{c}$ as a function of doping, but it also correlates with the maximum possible $T_\mathrm{c}$. While not all details are understood, the new scenario resembles what is now well established in terms of the sharing of charge between planar Cu and O. Here, the family dependence is predominantly set by the sharing of the parent hole between Cu and O (i.e.\@ by the charge transfer gap), which also determines the properties of the ICs. Rather independent of this, doping decreases the size of a temperature-independent pseudogap, in a universal density of states. While the magnetic response of the PCs saturates, that of the ICs increases with doping even after the pseudogap is fully closed. The condensation behavior of both carriers is linked, but differs distinctly. Clearly, both components are essential to cuprate superconductivity. It remains to be seen whether the described phenomenology can flow from a single fluid of correlated electrons, or whether two bands are necessary.
 
\vspace{0.3cm}\par\medskip

\begin{acknowledgements}
We acknowledge discussions with Boris Fine, Andreas Poeppl, and help from Crina Berbecariu (Leipzig) with finalisation of the manuscript, and financial support from Leipzig University. 
\end{acknowledgements}

{\flushleft \bf Author contributions\\}
D.B., J.H. contributed nearly equally to data analysis and led the preparation of the manuscript, J.N. was involved in early data analysis, A.L. was involved in data analysis and writing the manuscript, D.B. and S.T. contributed new measurements on \lsco.

{\flushleft \bf Competing Interests\\}
{ There are no competing interests for any of the authors.}\par\medskip

{\flushleft \bf Data Availability Statement \\}
All used data will be made available which enables reproduction of the results.\par\medskip

{\flushleft \bf Funding\\}
Funding of the research came from Leipzig University.\par\medskip
\vspace{0.5cm}

\bibliography{JH-CuprateJHc.bib}   

\begin{thebibliography}{46}%
\makeatletter
\providecommand \@ifxundefined [1]{%
 \@ifx{#1\undefined}
}%
\providecommand \@ifnum [1]{%
 \ifnum #1\expandafter \@firstoftwo
 \else \expandafter \@secondoftwo
 \fi
}%
\providecommand \@ifx [1]{%
 \ifx #1\expandafter \@firstoftwo
 \else \expandafter \@secondoftwo
 \fi
}%
\providecommand \natexlab [1]{#1}%
\providecommand \enquote  [1]{``#1''}%
\providecommand \bibnamefont  [1]{#1}%
\providecommand \bibfnamefont [1]{#1}%
\providecommand \citenamefont [1]{#1}%
\providecommand \href@noop [0]{\@secondoftwo}%
\providecommand \href [0]{\begingroup \@sanitize@url \@href}%
\providecommand \@href[1]{\@@startlink{#1}\@@href}%
\providecommand \@@href[1]{\endgroup#1\@@endlink}%
\providecommand \@sanitize@url [0]{\catcode `\\12\catcode `\$12\catcode
  `\&12\catcode `\#12\catcode `\^12\catcode `\_12\catcode `\%12\relax}%
\providecommand \@@startlink[1]{}%
\providecommand \@@endlink[0]{}%
\providecommand \url  [0]{\begingroup\@sanitize@url \@url }%
\providecommand \@url [1]{\endgroup\@href {#1}{\urlprefix }}%
\providecommand \urlprefix  [0]{URL }%
\providecommand \Eprint [0]{\href }%
\providecommand \doibase [0]{https://doi.org/}%
\providecommand \selectlanguage [0]{\@gobble}%
\providecommand \bibinfo  [0]{\@secondoftwo}%
\providecommand \bibfield  [0]{\@secondoftwo}%
\providecommand \translation [1]{[#1]}%
\providecommand \BibitemOpen [0]{}%
\providecommand \bibitemStop [0]{}%
\providecommand \bibitemNoStop [0]{.\EOS\space}%
\providecommand \EOS [0]{\spacefactor3000\relax}%
\providecommand \BibitemShut  [1]{\csname bibitem#1\endcsname}%
\let\auto@bib@innerbib\@empty
\bibitem [{\citenamefont {Bednorz}\ and\ \citenamefont
  {M{\"{u}}ller}(1986)}]{Bednorz1986}%
  \BibitemOpen
  \bibfield  {author} {\bibinfo {author} {\bibfnamefont {J.~G.}\ \bibnamefont
  {Bednorz}}\ and\ \bibinfo {author} {\bibfnamefont {K.~A.}\ \bibnamefont
  {M{\"{u}}ller}},\ }\bibfield  {title} {\bibinfo {title} {{Possible High
  $T_\mathrm{c}$ Superconductivity in the Ba-La-Cu-O System}},\ }\href
  {https://doi.org/10.1007/BF01303701} {\bibfield  {journal} {\bibinfo
  {journal} {Z. Phys. B Condens. Matter}\ }\textbf {\bibinfo {volume} {193}},\
  \bibinfo {pages} {189} (\bibinfo {year} {1986})}\BibitemShut {NoStop}%
\bibitem [{\citenamefont {Uemura}\ \emph {et~al.}(1989)\citenamefont {Uemura},
  \citenamefont {Luke}, \citenamefont {Sternlieb}, \citenamefont {Brewer},
  \citenamefont {Carolan}, \citenamefont {Hardy}, \citenamefont {Kadono},
  \citenamefont {Kempton}, \citenamefont {Kiefl}, \citenamefont {Kreitzman},
  \citenamefont {Mulhern}, \citenamefont {Riseman}, \citenamefont {Williams},
  \citenamefont {Yang}, \citenamefont {Uchida}, \citenamefont {Takagi},
  \citenamefont {Gopalakrishnan}, \citenamefont {Sleight}, \citenamefont
  {Subramanian}, \citenamefont {Chien}, \citenamefont {Cieplak}, \citenamefont
  {Xiao}, \citenamefont {Lee}, \citenamefont {Statt}, \citenamefont {Stronach},
  \citenamefont {Kossler},\ and\ \citenamefont {Yu}}]{Uemura1989}%
  \BibitemOpen
  \bibfield  {author} {\bibinfo {author} {\bibfnamefont {Y.~J.}\ \bibnamefont
  {Uemura}}, \bibinfo {author} {\bibfnamefont {G.~M.}\ \bibnamefont {Luke}},
  \bibinfo {author} {\bibfnamefont {B.~J.}\ \bibnamefont {Sternlieb}}, \bibinfo
  {author} {\bibfnamefont {J.~H.}\ \bibnamefont {Brewer}}, \bibinfo {author}
  {\bibfnamefont {J.~F.}\ \bibnamefont {Carolan}}, \bibinfo {author}
  {\bibfnamefont {W.~N.}\ \bibnamefont {Hardy}}, \bibinfo {author}
  {\bibfnamefont {R.}~\bibnamefont {Kadono}}, \bibinfo {author} {\bibfnamefont
  {J.~R.}\ \bibnamefont {Kempton}}, \bibinfo {author} {\bibfnamefont {R.~F.}\
  \bibnamefont {Kiefl}}, \bibinfo {author} {\bibfnamefont {S.~R.}\ \bibnamefont
  {Kreitzman}}, \bibinfo {author} {\bibfnamefont {P.}~\bibnamefont {Mulhern}},
  \bibinfo {author} {\bibfnamefont {T.~M.}\ \bibnamefont {Riseman}}, \bibinfo
  {author} {\bibfnamefont {D.~L.}\ \bibnamefont {Williams}}, \bibinfo {author}
  {\bibfnamefont {B.~X.}\ \bibnamefont {Yang}}, \bibinfo {author}
  {\bibfnamefont {S.}~\bibnamefont {Uchida}}, \bibinfo {author} {\bibfnamefont
  {H.}~\bibnamefont {Takagi}}, \bibinfo {author} {\bibfnamefont
  {J.}~\bibnamefont {Gopalakrishnan}}, \bibinfo {author} {\bibfnamefont
  {A.~W.}\ \bibnamefont {Sleight}}, \bibinfo {author} {\bibfnamefont {M.~A.}\
  \bibnamefont {Subramanian}}, \bibinfo {author} {\bibfnamefont {C.~L.}\
  \bibnamefont {Chien}}, \bibinfo {author} {\bibfnamefont {M.~Z.}\ \bibnamefont
  {Cieplak}}, \bibinfo {author} {\bibfnamefont {G.}~\bibnamefont {Xiao}},
  \bibinfo {author} {\bibfnamefont {V.~Y.}\ \bibnamefont {Lee}}, \bibinfo
  {author} {\bibfnamefont {B.~W.}\ \bibnamefont {Statt}}, \bibinfo {author}
  {\bibfnamefont {C.~E.}\ \bibnamefont {Stronach}}, \bibinfo {author}
  {\bibfnamefont {W.~J.}\ \bibnamefont {Kossler}},\ and\ \bibinfo {author}
  {\bibfnamefont {X.~H.}\ \bibnamefont {Yu}},\ }\bibfield  {title} {\bibinfo
  {title} {Universal correlations between {$T_c$} and $n_s/m^*$ (carrier
  density over effective mass) in high-{$T_c$} cuprate superconductors},\
  }\href {https://doi.org/10.1103/PhysRevLett.62.2317} {\bibfield  {journal}
  {\bibinfo  {journal} {Phys. Rev. Lett.}\ }\textbf {\bibinfo {volume} {62}},\
  \bibinfo {pages} {2317} (\bibinfo {year} {1989})}\BibitemShut {NoStop}%
\bibitem [{\citenamefont {Ohta}\ \emph {et~al.}(1991)\citenamefont {Ohta},
  \citenamefont {Tohyama},\ and\ \citenamefont {Maekawa}}]{Ohta1991}%
  \BibitemOpen
  \bibfield  {author} {\bibinfo {author} {\bibfnamefont {Y.}~\bibnamefont
  {Ohta}}, \bibinfo {author} {\bibfnamefont {T.}~\bibnamefont {Tohyama}},\ and\
  \bibinfo {author} {\bibfnamefont {S.}~\bibnamefont {Maekawa}},\ }\bibfield
  {title} {\bibinfo {title} {Apex oxygen and critical temperature in copper
  oxide superconductors: Universal correlation with the stability of local
  singlets},\ }\href {https://doi.org/10.1103/PhysRevB.43.2968} {\bibfield
  {journal} {\bibinfo  {journal} {Phys. Rev. B}\ }\textbf {\bibinfo {volume}
  {43}},\ \bibinfo {pages} {2968} (\bibinfo {year} {1991})}\BibitemShut
  {NoStop}%
\bibitem [{\citenamefont {Pavarini}\ \emph {et~al.}(2001)\citenamefont
  {Pavarini}, \citenamefont {Dasgupta}, \citenamefont {Saha-Dasgupta},
  \citenamefont {Jepsen},\ and\ \citenamefont {Andersen}}]{Pavarini2001}%
  \BibitemOpen
  \bibfield  {author} {\bibinfo {author} {\bibfnamefont {E.}~\bibnamefont
  {Pavarini}}, \bibinfo {author} {\bibfnamefont {I.}~\bibnamefont {Dasgupta}},
  \bibinfo {author} {\bibfnamefont {T.}~\bibnamefont {Saha-Dasgupta}}, \bibinfo
  {author} {\bibfnamefont {O.}~\bibnamefont {Jepsen}},\ and\ \bibinfo {author}
  {\bibfnamefont {O.~K.}\ \bibnamefont {Andersen}},\ }\bibfield  {title}
  {\bibinfo {title} {Band-structure trend in hole-doped cuprates and
  correlation with {$T_\mathrm{c, max}$}},\ }\href
  {https://doi.org/10.1103/PhysRevLett.87.047003} {\bibfield  {journal}
  {\bibinfo  {journal} {Phys. Rev. Lett.}\ }\textbf {\bibinfo {volume} {87}},\
  \bibinfo {pages} {047003} (\bibinfo {year} {2001})}\BibitemShut {NoStop}%
\bibitem [{\citenamefont {O'Mahony}\ \emph {et~al.}(2022)\citenamefont
  {O'Mahony}, \citenamefont {Ren}, \citenamefont {Chen}, \citenamefont {Chong},
  \citenamefont {Liu}, \citenamefont {Eisaki}, \citenamefont {Uchida},
  \citenamefont {Hamidian},\ and\ \citenamefont {Davis}}]{Mahony2022}%
  \BibitemOpen
  \bibfield  {author} {\bibinfo {author} {\bibfnamefont {S.~M.}\ \bibnamefont
  {O'Mahony}}, \bibinfo {author} {\bibfnamefont {W.}~\bibnamefont {Ren}},
  \bibinfo {author} {\bibfnamefont {W.}~\bibnamefont {Chen}}, \bibinfo {author}
  {\bibfnamefont {Y.~X.}\ \bibnamefont {Chong}}, \bibinfo {author}
  {\bibfnamefont {X.}~\bibnamefont {Liu}}, \bibinfo {author} {\bibfnamefont
  {H.}~\bibnamefont {Eisaki}}, \bibinfo {author} {\bibfnamefont
  {S.}~\bibnamefont {Uchida}}, \bibinfo {author} {\bibfnamefont {M.~H.}\
  \bibnamefont {Hamidian}},\ and\ \bibinfo {author} {\bibfnamefont {J.~C.}\
  \bibnamefont {Davis}},\ }\bibfield  {title} {\bibinfo {title} {On the
  electron pairing mechanism of copper-oxide high temperature
  superconductors},\ }\bibfield  {journal} {\bibinfo  {journal} {PNAS}\
  }\textbf {\bibinfo {volume} {119}},\ \href
  {https://doi.org/10.1073/pnas.2207449119} {10.1073/pnas.2207449119} (\bibinfo
  {year} {2022})\BibitemShut {NoStop}%
\bibitem [{\citenamefont {Zheng}\ \emph {et~al.}(1995)\citenamefont {Zheng},
  \citenamefont {Kitaoka}, \citenamefont {Ishida},\ and\ \citenamefont
  {Asayama}}]{Zheng1995h}%
  \BibitemOpen
  \bibfield  {author} {\bibinfo {author} {\bibfnamefont {G.-q.}\ \bibnamefont
  {Zheng}}, \bibinfo {author} {\bibfnamefont {Y.}~\bibnamefont {Kitaoka}},
  \bibinfo {author} {\bibfnamefont {K.}~\bibnamefont {Ishida}},\ and\ \bibinfo
  {author} {\bibfnamefont {K.}~\bibnamefont {Asayama}},\ }\bibfield  {title}
  {\bibinfo {title} {Local {Hole} {Distribution} in the {CuO$_2$} {Plane} of
  {High}-{T$_\mathrm{c}$} {Cu}-{Oxides} {Studied} by {Cu} and {Oxygen}
  {NQR}/{NMR}},\ }\href {https://doi.org/10.1143/JPSJ.64.2524} {\bibfield
  {journal} {\bibinfo  {journal} {J. Phys. Soc. Jpn.}\ }\textbf {\bibinfo
  {volume} {64}},\ \bibinfo {pages} {2524} (\bibinfo {year}
  {1995})}\BibitemShut {NoStop}%
\bibitem [{\citenamefont {Haase}\ \emph {et~al.}(2004)\citenamefont {Haase},
  \citenamefont {Sushkov}, \citenamefont {Horsch},\ and\ \citenamefont
  {Williams}}]{Haase2004}%
  \BibitemOpen
  \bibfield  {author} {\bibinfo {author} {\bibfnamefont {J.}~\bibnamefont
  {Haase}}, \bibinfo {author} {\bibfnamefont {O.~P.}\ \bibnamefont {Sushkov}},
  \bibinfo {author} {\bibfnamefont {P.}~\bibnamefont {Horsch}},\ and\ \bibinfo
  {author} {\bibfnamefont {G.~V.~M.}\ \bibnamefont {Williams}},\ }\bibfield
  {title} {\bibinfo {title} {{Planar Cu and O hole densities in
  high-$T_\text{c}$ cuprates determined with NMR}},\ }\href
  {https://doi.org/10.1103/PhysRevB.69.094504} {\bibfield  {journal} {\bibinfo
  {journal} {Phys. Rev. B}\ }\textbf {\bibinfo {volume} {69}},\ \bibinfo
  {pages} {94504} (\bibinfo {year} {2004})}\BibitemShut {NoStop}%
\bibitem [{\citenamefont {Rybicki}\ \emph {et~al.}(2016)\citenamefont
  {Rybicki}, \citenamefont {Jurkutat}, \citenamefont {Reichardt}, \citenamefont
  {Kapusta},\ and\ \citenamefont {Haase}}]{Rybicki2016}%
  \BibitemOpen
  \bibfield  {author} {\bibinfo {author} {\bibfnamefont {D.}~\bibnamefont
  {Rybicki}}, \bibinfo {author} {\bibfnamefont {M.}~\bibnamefont {Jurkutat}},
  \bibinfo {author} {\bibfnamefont {S.}~\bibnamefont {Reichardt}}, \bibinfo
  {author} {\bibfnamefont {C.}~\bibnamefont {Kapusta}},\ and\ \bibinfo {author}
  {\bibfnamefont {J.}~\bibnamefont {Haase}},\ }\bibfield  {title} {\bibinfo
  {title} {{Perspective on the phase diagram of cuprate high-temperature
  superconductors}},\ }\href {https://doi.org/10.1038/ncomms11413} {\bibfield
  {journal} {\bibinfo  {journal} {Nat. Commun.}\ }\textbf {\bibinfo {volume}
  {7}},\ \bibinfo {pages} {1} (\bibinfo {year} {2016})}\BibitemShut {NoStop}%
\bibitem [{\citenamefont {Kowalski}\ \emph {et~al.}(2021)\citenamefont
  {Kowalski}, \citenamefont {Dash}, \citenamefont {S{\'e}mon}, \citenamefont
  {S{\'e}n{\'e}chal},\ and\ \citenamefont {Tremblay}}]{Kowalski2021}%
  \BibitemOpen
  \bibfield  {author} {\bibinfo {author} {\bibfnamefont {N.}~\bibnamefont
  {Kowalski}}, \bibinfo {author} {\bibfnamefont {S.~S.}\ \bibnamefont {Dash}},
  \bibinfo {author} {\bibfnamefont {P.}~\bibnamefont {S{\'e}mon}}, \bibinfo
  {author} {\bibfnamefont {D.}~\bibnamefont {S{\'e}n{\'e}chal}},\ and\ \bibinfo
  {author} {\bibfnamefont {A.-M.}\ \bibnamefont {Tremblay}},\ }\bibfield
  {title} {\bibinfo {title} {Oxygen hole content, charge-transfer gap,
  covalency, and cuprate superconductivity},\ }\href
  {https://doi.org/10.1073/pnas.2106476118} {\bibfield  {journal} {\bibinfo
  {journal} {PNAS}\ }\textbf {\bibinfo {volume} {118}},\ \bibinfo {pages} {1}
  (\bibinfo {year} {2021})}\BibitemShut {NoStop}%
\bibitem [{\citenamefont {Bacq-Labreuil}\ \emph {et~al.}(2025)\citenamefont
  {Bacq-Labreuil}, \citenamefont {Lacasse}, \citenamefont {Tremblay},
  \citenamefont {S{\'e}n{\'e}chal},\ and\ \citenamefont {Haule}}]{Bacq2025}%
  \BibitemOpen
  \bibfield  {author} {\bibinfo {author} {\bibfnamefont {B.}~\bibnamefont
  {Bacq-Labreuil}}, \bibinfo {author} {\bibfnamefont {B.}~\bibnamefont
  {Lacasse}}, \bibinfo {author} {\bibfnamefont {A.-M.~S.}\ \bibnamefont
  {Tremblay}}, \bibinfo {author} {\bibfnamefont {D.}~\bibnamefont
  {S{\'e}n{\'e}chal}},\ and\ \bibinfo {author} {\bibfnamefont {K.}~\bibnamefont
  {Haule}},\ }\bibfield  {title} {\bibinfo {title} {Toward an {Ab} {Initio}
  {Theory} of {High}-{Temperature} {Superconductors}: {A} {Study} of
  {Multilayer} {Cuprates}},\ }\href
  {https://doi.org/10.1103/PhysRevX.15.021071} {\bibfield  {journal} {\bibinfo
  {journal} {Phys. Rev. X}\ }\textbf {\bibinfo {volume} {15}},\ \bibinfo
  {pages} {021071} (\bibinfo {year} {2025})}\BibitemShut {NoStop}%
\bibitem [{\citenamefont {Jurkutat}\ \emph {et~al.}(2023)\citenamefont
  {Jurkutat}, \citenamefont {Kattinger}, \citenamefont {Tsankov}, \citenamefont
  {Reznicek}, \citenamefont {Erb},\ and\ \citenamefont {Haase}}]{Jurkutat2023}%
  \BibitemOpen
  \bibfield  {author} {\bibinfo {author} {\bibfnamefont {M.}~\bibnamefont
  {Jurkutat}}, \bibinfo {author} {\bibfnamefont {C.}~\bibnamefont {Kattinger}},
  \bibinfo {author} {\bibfnamefont {S.}~\bibnamefont {Tsankov}}, \bibinfo
  {author} {\bibfnamefont {R.}~\bibnamefont {Reznicek}}, \bibinfo {author}
  {\bibfnamefont {A.}~\bibnamefont {Erb}},\ and\ \bibinfo {author}
  {\bibfnamefont {J.}~\bibnamefont {Haase}},\ }\bibfield  {title} {\bibinfo
  {title} {How pressure enhances the critical temperature for high temperature
  superconductivity in {YBa}$_2${Cu}$_3${O}$_{6+y}$},\ }\href
  {https://doi.org/10.1073/pnas.2215458120} {\bibfield  {journal} {\bibinfo
  {journal} {PNAS}\ }\textbf {\bibinfo {volume} {120}},\ \bibinfo {pages}
  {e2215458120} (\bibinfo {year} {2023})}\BibitemShut {NoStop}%
\bibitem [{\citenamefont {Slichter}(2007)}]{Slichter2007}%
  \BibitemOpen
  \bibfield  {author} {\bibinfo {author} {\bibfnamefont {C.~P.}\ \bibnamefont
  {Slichter}},\ }\bibfield  {title} {\bibinfo {title} {{Magnetic Resonance
  Studies of High Temperature Superconductors}},\ }in\ \href
  {https://doi.org/10.1007/978-0-387-68734-6\_5} {\emph {\bibinfo {booktitle}
  {Handbook of High-Temperature Superconductivity}}},\ \bibinfo {editor}
  {edited by\ \bibinfo {editor} {\bibfnamefont {J.~R.}\ \bibnamefont
  {Schrieffer}}\ and\ \bibinfo {editor} {\bibfnamefont {J.~S.}\ \bibnamefont
  {Brooks}}}\ (\bibinfo  {publisher} {Springer},\ \bibinfo {address} {New
  York},\ \bibinfo {year} {2007})\ pp.\ \bibinfo {pages} {215--256}\BibitemShut
  {NoStop}%
\bibitem [{\citenamefont {Yosida}(1958)}]{Yosida1958}%
  \BibitemOpen
  \bibfield  {author} {\bibinfo {author} {\bibfnamefont {K.}~\bibnamefont
  {Yosida}},\ }\bibfield  {title} {\bibinfo {title} {{Paramagnetic
  Susceptibility in Superconductors}},\ }\href
  {https://doi.org/10.1103/PhysRev.110.769} {\bibfield  {journal} {\bibinfo
  {journal} {Phys. Rev.}\ }\textbf {\bibinfo {volume} {110}},\ \bibinfo {pages}
  {769} (\bibinfo {year} {1958})}\BibitemShut {NoStop}%
\bibitem [{\citenamefont {Schrieffer}(2018)}]{Schrieffer2018}%
  \BibitemOpen
  \bibfield  {author} {\bibinfo {author} {\bibfnamefont {J.~R.}\ \bibnamefont
  {Schrieffer}},\ }\href@noop {} {\emph {\bibinfo {title} {Theory of
  superconductivity}}}\ (\bibinfo  {publisher} {CRC press},\ \bibinfo {year}
  {2018})\BibitemShut {NoStop}%
\bibitem [{\citenamefont {Carter}\ \emph {et~al.}(1977)\citenamefont {Carter},
  \citenamefont {Bennett},\ and\ \citenamefont {Kahan}}]{Carter1976}%
  \BibitemOpen
  \bibfield  {author} {\bibinfo {author} {\bibfnamefont {G.~S.}\ \bibnamefont
  {Carter}}, \bibinfo {author} {\bibfnamefont {L.~H.}\ \bibnamefont
  {Bennett}},\ and\ \bibinfo {author} {\bibfnamefont {D.~J.}\ \bibnamefont
  {Kahan}},\ }\href@noop {} {\emph {\bibinfo {title} {{Metallic Shifts in NMR,
  A review of theory and comprehensive critical data compilation of metallic
  materials}}}},\ edited by\ \bibinfo {editor} {\bibfnamefont {T.~B.~M.}\
  \bibnamefont {B.~Chalmers}, \bibfnamefont {J.~W.~Christian}},\ Vol.~\bibinfo
  {volume} {20}\ (\bibinfo  {publisher} {Pergamon Press, Oxford, New York,
  Toronto, Sidney, Paris, Frankfurt},\ \bibinfo {year} {1977})\BibitemShut
  {NoStop}%
\bibitem [{\citenamefont {Haase}\ \emph {et~al.}(2017)\citenamefont {Haase},
  \citenamefont {Jurkutat},\ and\ \citenamefont {Kohlrautz}}]{Haase2017}%
  \BibitemOpen
  \bibfield  {author} {\bibinfo {author} {\bibfnamefont {J.}~\bibnamefont
  {Haase}}, \bibinfo {author} {\bibfnamefont {M.}~\bibnamefont {Jurkutat}},\
  and\ \bibinfo {author} {\bibfnamefont {J.}~\bibnamefont {Kohlrautz}},\
  }\bibfield  {title} {\bibinfo {title} {{Contrasting Phenomenology of NMR
  Shifts in Cuprate Superconductors}},\ }\href
  {https://doi.org/10.3390/condmat2020016} {\bibfield  {journal} {\bibinfo
  {journal} {Condens. Matter}\ }\textbf {\bibinfo {volume} {2}},\ \bibinfo
  {pages} {16} (\bibinfo {year} {2017})}\BibitemShut {NoStop}%
\bibitem [{\citenamefont {Takigawa}\ \emph {et~al.}(1991)\citenamefont
  {Takigawa}, \citenamefont {Reyes}, \citenamefont {Hammel}, \citenamefont
  {Thompson}, \citenamefont {Heffner}, \citenamefont {Fisk},\ and\
  \citenamefont {Ott}}]{Takigawa1991}%
  \BibitemOpen
  \bibfield  {author} {\bibinfo {author} {\bibfnamefont {M.}~\bibnamefont
  {Takigawa}}, \bibinfo {author} {\bibfnamefont {A.~P.}\ \bibnamefont {Reyes}},
  \bibinfo {author} {\bibfnamefont {P.~C.}\ \bibnamefont {Hammel}}, \bibinfo
  {author} {\bibfnamefont {J.~D.}\ \bibnamefont {Thompson}}, \bibinfo {author}
  {\bibfnamefont {R.~H.}\ \bibnamefont {Heffner}}, \bibinfo {author}
  {\bibfnamefont {Z.}~\bibnamefont {Fisk}},\ and\ \bibinfo {author}
  {\bibfnamefont {K.~C.}\ \bibnamefont {Ott}},\ }\bibfield  {title} {\bibinfo
  {title} {{Cu and O NMR studies of the magnetic properties of
  YBa$_2$Cu$_3$O$_{6.63}$ ($T_\mathrm{c}$={62}{K})}},\ }\href
  {https://doi.org/10.1103/PhysRevB.43.247} {\bibfield  {journal} {\bibinfo
  {journal} {Phys. Rev. B}\ }\textbf {\bibinfo {volume} {43}},\ \bibinfo
  {pages} {247} (\bibinfo {year} {1991})}\BibitemShut {NoStop}%
\bibitem [{\citenamefont {Bankay}\ \emph {et~al.}(1994)\citenamefont {Bankay},
  \citenamefont {Mali}, \citenamefont {Roos},\ and\ \citenamefont
  {Brinkmann}}]{Bankay1994}%
  \BibitemOpen
  \bibfield  {author} {\bibinfo {author} {\bibfnamefont {M.}~\bibnamefont
  {Bankay}}, \bibinfo {author} {\bibfnamefont {M.}~\bibnamefont {Mali}},
  \bibinfo {author} {\bibfnamefont {J.}~\bibnamefont {Roos}},\ and\ \bibinfo
  {author} {\bibfnamefont {D.}~\bibnamefont {Brinkmann}},\ }\bibfield  {title}
  {\bibinfo {title} {{Single-spin fluid, spin gap, and d-wave pairing in
  YBa$_2$Cu$_4$O$_8$: A NMR and NQR study}},\ }\href
  {https://doi.org/10.1103/PhysRevB.50.6416} {\bibfield  {journal} {\bibinfo
  {journal} {Phys. Rev. B}\ }\textbf {\bibinfo {volume} {50}},\ \bibinfo
  {pages} {6416} (\bibinfo {year} {1994})}\BibitemShut {NoStop}%
\bibitem [{\citenamefont {Kitaoka}\ \emph {et~al.}(1991)\citenamefont
  {Kitaoka}, \citenamefont {Fujiwara}, \citenamefont {Ishida}, \citenamefont
  {Asayama}, \citenamefont {Shimakawa}, \citenamefont {Manako},\ and\
  \citenamefont {Kubo}}]{Kitaoka1991}%
  \BibitemOpen
  \bibfield  {author} {\bibinfo {author} {\bibfnamefont {Y.}~\bibnamefont
  {Kitaoka}}, \bibinfo {author} {\bibfnamefont {K.}~\bibnamefont {Fujiwara}},
  \bibinfo {author} {\bibfnamefont {K.}~\bibnamefont {Ishida}}, \bibinfo
  {author} {\bibfnamefont {K.}~\bibnamefont {Asayama}}, \bibinfo {author}
  {\bibfnamefont {Y.}~\bibnamefont {Shimakawa}}, \bibinfo {author}
  {\bibfnamefont {T.}~\bibnamefont {Manako}},\ and\ \bibinfo {author}
  {\bibfnamefont {Y.}~\bibnamefont {Kubo}},\ }\bibfield  {title} {\bibinfo
  {title} {Spin dynamics in heavily-doped high-{$T_c$} superconductors
  {Tl$_2$Ba$_2$CuO$_{6+y}$} with a single {CuO$_2$} layer studied by
  {$^{63}$Cu} and {$^{205}$Tl} {NMR}},\ }\href
  {https://doi.org/https://doi.org/10.1016/0921-4534(91)90018-T} {\bibfield
  {journal} {\bibinfo  {journal} {Physica C: Supercond.}\ }\textbf {\bibinfo
  {volume} {179}},\ \bibinfo {pages} {107} (\bibinfo {year}
  {1991})}\BibitemShut {NoStop}%
\bibitem [{\citenamefont {Barrett}\ \emph {et~al.}(1990)\citenamefont
  {Barrett}, \citenamefont {Durand}, \citenamefont {Pennington}, \citenamefont
  {Slichter}, \citenamefont {Friedmann}, \citenamefont {Rice},\ and\
  \citenamefont {Ginsberg}}]{Barrett1990b}%
  \BibitemOpen
  \bibfield  {author} {\bibinfo {author} {\bibfnamefont {S.~E.}\ \bibnamefont
  {Barrett}}, \bibinfo {author} {\bibfnamefont {D.~J.}\ \bibnamefont {Durand}},
  \bibinfo {author} {\bibfnamefont {C.~H.}\ \bibnamefont {Pennington}},
  \bibinfo {author} {\bibfnamefont {C.~P.}\ \bibnamefont {Slichter}}, \bibinfo
  {author} {\bibfnamefont {T.~A.}\ \bibnamefont {Friedmann}}, \bibinfo {author}
  {\bibfnamefont {J.~P.}\ \bibnamefont {Rice}},\ and\ \bibinfo {author}
  {\bibfnamefont {D.~M.}\ \bibnamefont {Ginsberg}},\ }\bibfield  {title}
  {\bibinfo {title} {{$^{63}$Cu Knight shifts in the superconducting state of
  {YBa$_2$Cu$_3$O$_{7-\delta}$} ($T_\mathrm{c}=90$K)}},\ }\href
  {https://doi.org/10.1103/PhysRevB.41.6283} {\bibfield  {journal} {\bibinfo
  {journal} {Phys. Rev. B}\ }\textbf {\bibinfo {volume} {41}},\ \bibinfo
  {pages} {6283} (\bibinfo {year} {1990})}\BibitemShut {NoStop}%
\bibitem [{\citenamefont {Haase}\ \emph {et~al.}(2009)\citenamefont {Haase},
  \citenamefont {Slichter},\ and\ \citenamefont {Williams}}]{Haase2009b}%
  \BibitemOpen
  \bibfield  {author} {\bibinfo {author} {\bibfnamefont {J.}~\bibnamefont
  {Haase}}, \bibinfo {author} {\bibfnamefont {C.~P.}\ \bibnamefont
  {Slichter}},\ and\ \bibinfo {author} {\bibfnamefont {G.~V.~M.}\ \bibnamefont
  {Williams}},\ }\bibfield  {title} {\bibinfo {title} {{Evidence for two
  electronic components in high-temperature superconductivity from NMR}},\
  }\href {https://doi.org/10.1088/0953-8984/21/45/455702} {\bibfield  {journal}
  {\bibinfo  {journal} {J. Phys. Condens. Matter}\ }\textbf {\bibinfo {volume}
  {21}},\ \bibinfo {pages} {455702} (\bibinfo {year} {2009})}\BibitemShut
  {NoStop}%
\bibitem [{\citenamefont {Haase}\ \emph {et~al.}(2012)\citenamefont {Haase},
  \citenamefont {Rybicki}, \citenamefont {Slichter}, \citenamefont {Greven},
  \citenamefont {Yu}, \citenamefont {Li},\ and\ \citenamefont
  {Zhao}}]{Haase2012}%
  \BibitemOpen
  \bibfield  {author} {\bibinfo {author} {\bibfnamefont {J.}~\bibnamefont
  {Haase}}, \bibinfo {author} {\bibfnamefont {D.}~\bibnamefont {Rybicki}},
  \bibinfo {author} {\bibfnamefont {C.~P.}\ \bibnamefont {Slichter}}, \bibinfo
  {author} {\bibfnamefont {M.}~\bibnamefont {Greven}}, \bibinfo {author}
  {\bibfnamefont {G.}~\bibnamefont {Yu}}, \bibinfo {author} {\bibfnamefont
  {Y.}~\bibnamefont {Li}},\ and\ \bibinfo {author} {\bibfnamefont
  {X.}~\bibnamefont {Zhao}},\ }\bibfield  {title} {\bibinfo {title}
  {{Two-component uniform spin susceptibility of superconducting
  \ce{HgBa_2CuO_{4+\delta}} single crystals measured using \ce{{}^{63}Cu} and
  \ce{{}^{199}Hg} nuclear magnetic resonance}},\ }\href
  {https://doi.org/10.1103/PhysRevB.85.104517} {\bibfield  {journal} {\bibinfo
  {journal} {Phys. Rev. B}\ }\textbf {\bibinfo {volume} {85}},\ \bibinfo
  {pages} {104517} (\bibinfo {year} {2012})}\BibitemShut {NoStop}%
\bibitem [{\citenamefont {Rybicki}\ \emph {et~al.}(2015)\citenamefont
  {Rybicki}, \citenamefont {Kohlrautz}, \citenamefont {Haase}, \citenamefont
  {Greven}, \citenamefont {Zhao}, \citenamefont {Chan}, \citenamefont {Dorow},\
  and\ \citenamefont {Veit}}]{Rybicki2015}%
  \BibitemOpen
  \bibfield  {author} {\bibinfo {author} {\bibfnamefont {D.}~\bibnamefont
  {Rybicki}}, \bibinfo {author} {\bibfnamefont {J.}~\bibnamefont {Kohlrautz}},
  \bibinfo {author} {\bibfnamefont {J.}~\bibnamefont {Haase}}, \bibinfo
  {author} {\bibfnamefont {M.}~\bibnamefont {Greven}}, \bibinfo {author}
  {\bibfnamefont {X.}~\bibnamefont {Zhao}}, \bibinfo {author} {\bibfnamefont
  {M.~K.}\ \bibnamefont {Chan}}, \bibinfo {author} {\bibfnamefont {C.~J.}\
  \bibnamefont {Dorow}},\ and\ \bibinfo {author} {\bibfnamefont {M.~J.}\
  \bibnamefont {Veit}},\ }\bibfield  {title} {\bibinfo {title} {{Electronic
  spin susceptibilities and superconductivity in HgBa$_2$CuO$_{4+\delta}$ from
  nuclear magnetic resonance}},\ }\href
  {https://doi.org/10.1103/PhysRevB.92.081115} {\bibfield  {journal} {\bibinfo
  {journal} {Phys. Rev. B}\ }\textbf {\bibinfo {volume} {92}},\ \bibinfo
  {pages} {081115(R)} (\bibinfo {year} {2015})}\BibitemShut {NoStop}%
\bibitem [{\citenamefont {Nachtigal}\ \emph {et~al.}(2020)\citenamefont
  {Nachtigal}, \citenamefont {Avramovska}, \citenamefont {Erb}, \citenamefont
  {Pavi{\'c}evi{\'c}}, \citenamefont {Guehne},\ and\ \citenamefont
  {Haase}}]{Nachtigal2020}%
  \BibitemOpen
  \bibfield  {author} {\bibinfo {author} {\bibfnamefont {J.}~\bibnamefont
  {Nachtigal}}, \bibinfo {author} {\bibfnamefont {M.}~\bibnamefont
  {Avramovska}}, \bibinfo {author} {\bibfnamefont {A.}~\bibnamefont {Erb}},
  \bibinfo {author} {\bibfnamefont {D.}~\bibnamefont {Pavi{\'c}evi{\'c}}},
  \bibinfo {author} {\bibfnamefont {R.}~\bibnamefont {Guehne}},\ and\ \bibinfo
  {author} {\bibfnamefont {J.}~\bibnamefont {Haase}},\ }\bibfield  {title}
  {\bibinfo {title} {{Temperature-Independent Cuprate Pseudogap from Planar
  Oxygen NMR}},\ }\href {https://doi.org/10.3390/condmat5040066} {\bibfield
  {journal} {\bibinfo  {journal} {Condens. Matter}\ }\textbf {\bibinfo {volume}
  {5}},\ \bibinfo {pages} {66} (\bibinfo {year} {2020})}\BibitemShut {NoStop}%
\bibitem [{\citenamefont {Avramovska}\ \emph
  {et~al.}(2022{\natexlab{a}})\citenamefont {Avramovska}, \citenamefont
  {Nachtigal}, \citenamefont {Tsankov},\ and\ \citenamefont
  {Haase}}]{Avramovska2022}%
  \BibitemOpen
  \bibfield  {author} {\bibinfo {author} {\bibfnamefont {M.}~\bibnamefont
  {Avramovska}}, \bibinfo {author} {\bibfnamefont {J.}~\bibnamefont
  {Nachtigal}}, \bibinfo {author} {\bibfnamefont {S.}~\bibnamefont {Tsankov}},\
  and\ \bibinfo {author} {\bibfnamefont {J.}~\bibnamefont {Haase}},\ }\bibfield
   {title} {\bibinfo {title} {Planar {Cu} and {O} {NMR} and the {Pseudogap} of
  {Cuprate} {Superconductors}},\ }\href
  {https://doi.org/10.3390/condmat7010021} {\bibfield  {journal} {\bibinfo
  {journal} {Condens. Matter}\ }\textbf {\bibinfo {volume} {7}},\ \bibinfo
  {pages} {21} (\bibinfo {year} {2022}{\natexlab{a}})}\BibitemShut {NoStop}%
\bibitem [{\citenamefont {Avramovska}\ \emph
  {et~al.}(2022{\natexlab{b}})\citenamefont {Avramovska}, \citenamefont
  {Nachtigal},\ and\ \citenamefont {Haase}}]{Avramovska2022b}%
  \BibitemOpen
  \bibfield  {author} {\bibinfo {author} {\bibfnamefont {M.}~\bibnamefont
  {Avramovska}}, \bibinfo {author} {\bibfnamefont {J.}~\bibnamefont
  {Nachtigal}},\ and\ \bibinfo {author} {\bibfnamefont {J.}~\bibnamefont
  {Haase}},\ }\bibfield  {title} {\bibinfo {title} {{Temperature independent
  pseudogap from $^{17}$O and $^{89}$Y {NMR} and the single component
  picture}},\ }\href {https://doi.org/10.1007/s10948-022-06171-2} {\bibfield
  {journal} {\bibinfo  {journal} {J. Supercond. Nov. Magn.}\ }\textbf {\bibinfo
  {volume} {35}},\ \bibinfo {pages} {1761} (\bibinfo {year}
  {2022}{\natexlab{b}})}\BibitemShut {NoStop}%
\bibitem [{\citenamefont {Renold}\ \emph {et~al.}(2003)\citenamefont {Renold},
  \citenamefont {Heine}, \citenamefont {Weber},\ and\ \citenamefont
  {Meier}}]{Renold2003}%
  \BibitemOpen
  \bibfield  {author} {\bibinfo {author} {\bibfnamefont {S.}~\bibnamefont
  {Renold}}, \bibinfo {author} {\bibfnamefont {T.}~\bibnamefont {Heine}},
  \bibinfo {author} {\bibfnamefont {J.}~\bibnamefont {Weber}},\ and\ \bibinfo
  {author} {\bibfnamefont {P.~F.}\ \bibnamefont {Meier}},\ }\bibfield  {title}
  {\bibinfo {title} {{Nuclear magnetic resonance chemical shifts and
  paramagnetic field modifications in La$_{2}$CuO$_{4}$}},\ }\href
  {https://doi.org/10.1103/PhysRevB.67.024501} {\bibfield  {journal} {\bibinfo
  {journal} {Phys. Rev. B}\ }\textbf {\bibinfo {volume} {67}},\ \bibinfo
  {pages} {024501} (\bibinfo {year} {2003})}\BibitemShut {NoStop}%
\bibitem [{\citenamefont {H{\"u}sser}\ \emph {et~al.}(2000)\citenamefont
  {H{\"u}sser}, \citenamefont {Suter}, \citenamefont {Stoll},\ and\
  \citenamefont {Meier}}]{Huesser2000}%
  \BibitemOpen
  \bibfield  {author} {\bibinfo {author} {\bibfnamefont {P.}~\bibnamefont
  {H{\"u}sser}}, \bibinfo {author} {\bibfnamefont {H.~U.}\ \bibnamefont
  {Suter}}, \bibinfo {author} {\bibfnamefont {E.~P.}\ \bibnamefont {Stoll}},\
  and\ \bibinfo {author} {\bibfnamefont {P.~F.}\ \bibnamefont {Meier}},\
  }\bibfield  {title} {\bibinfo {title} {First-principles calculations of
  hyperfine interactions in {La}$_2${CuO}$_4$},\ }\href
  {https://doi.org/10.1103/PhysRevB.61.1567} {\bibfield  {journal} {\bibinfo
  {journal} {Phys. Rev. B}\ }\textbf {\bibinfo {volume} {61}},\ \bibinfo
  {pages} {1567} (\bibinfo {year} {2000})}\BibitemShut {NoStop}%
\bibitem [{\citenamefont {Pennington}\ \emph {et~al.}(1989)\citenamefont
  {Pennington}, \citenamefont {Durand}, \citenamefont {Slichter}, \citenamefont
  {Rice}, \citenamefont {Bukowski},\ and\ \citenamefont
  {Ginsberg}}]{Pennington1989}%
  \BibitemOpen
  \bibfield  {author} {\bibinfo {author} {\bibfnamefont {C.~H.}\ \bibnamefont
  {Pennington}}, \bibinfo {author} {\bibfnamefont {D.~J.}\ \bibnamefont
  {Durand}}, \bibinfo {author} {\bibfnamefont {C.~P.}\ \bibnamefont
  {Slichter}}, \bibinfo {author} {\bibfnamefont {J.~P.}\ \bibnamefont {Rice}},
  \bibinfo {author} {\bibfnamefont {E.~D.}\ \bibnamefont {Bukowski}},\ and\
  \bibinfo {author} {\bibfnamefont {D.~M.}\ \bibnamefont {Ginsberg}},\
  }\bibfield  {title} {\bibinfo {title} {{Static and dynamic Cu NMR tensors of
  YBa$_2$Cu$_3$O$_{7-\delta}$}},\ }\href
  {https://doi.org/10.1103/PhysRevB.39.2902} {\bibfield  {journal} {\bibinfo
  {journal} {Phys. Rev. B}\ }\textbf {\bibinfo {volume} {39}},\ \bibinfo
  {pages} {2902} (\bibinfo {year} {1989})}\BibitemShut {NoStop}%
\bibitem [{\citenamefont {Jurkutat}\ \emph {et~al.}(2014)\citenamefont
  {Jurkutat}, \citenamefont {Rybicki}, \citenamefont {Sushkov}, \citenamefont
  {Williams}, \citenamefont {Erb},\ and\ \citenamefont {Haase}}]{Jurkutat2014}%
  \BibitemOpen
  \bibfield  {author} {\bibinfo {author} {\bibfnamefont {M.}~\bibnamefont
  {Jurkutat}}, \bibinfo {author} {\bibfnamefont {D.}~\bibnamefont {Rybicki}},
  \bibinfo {author} {\bibfnamefont {O.~P.}\ \bibnamefont {Sushkov}}, \bibinfo
  {author} {\bibfnamefont {G.~V.~M.}\ \bibnamefont {Williams}}, \bibinfo
  {author} {\bibfnamefont {A.}~\bibnamefont {Erb}},\ and\ \bibinfo {author}
  {\bibfnamefont {J.}~\bibnamefont {Haase}},\ }\bibfield  {title} {\bibinfo
  {title} {{Distribution of electrons and holes in cuprate superconductors as
  determined from $^{17}$O and $^{63}$Cu nuclear magnetic resonance}},\ }\href
  {https://doi.org/10.1103/PhysRevB.90.140504} {\bibfield  {journal} {\bibinfo
  {journal} {Phys. Rev. B}\ }\textbf {\bibinfo {volume} {90}},\ \bibinfo
  {pages} {140504(R)} (\bibinfo {year} {2014})}\BibitemShut {NoStop}%
\bibitem [{\citenamefont {M{\"u}ller}(1995)}]{Mueller1995}%
  \BibitemOpen
  \bibfield  {author} {\bibinfo {author} {\bibfnamefont {K.~A.}\ \bibnamefont
  {M{\"u}ller}},\ }\bibfield  {title} {\bibinfo {title} {Possible coexistence
  of s- and d-wave condensates in copper oxide superconductors},\ }\href
  {https://doi.org/10.1038/377133a0} {\bibfield  {journal} {\bibinfo  {journal}
  {Nature}\ }\textbf {\bibinfo {volume} {377}},\ \bibinfo {pages} {133}
  (\bibinfo {year} {1995})}\BibitemShut {NoStop}%
\bibitem [{\citenamefont {Tallon}\ and\ \citenamefont
  {Storey}(2022)}]{Tallon2022}%
  \BibitemOpen
  \bibfield  {author} {\bibinfo {author} {\bibfnamefont {J.~L.}\ \bibnamefont
  {Tallon}}\ and\ \bibinfo {author} {\bibfnamefont {J.~G.}\ \bibnamefont
  {Storey}},\ }\bibfield  {title} {\bibinfo {title} {Thermodynamics of the
  pseudogap in cuprates},\ }\href {https://doi.org/10.3389/fphy.2022.1030616}
  {\bibfield  {journal} {\bibinfo  {journal} {Front. Phys.}\ }\textbf {\bibinfo
  {volume} {10}},\ \bibinfo {pages} {1030616} (\bibinfo {year}
  {2022})}\BibitemShut {NoStop}%
\bibitem [{\citenamefont {Loram}\ \emph {et~al.}(1998)\citenamefont {Loram},
  \citenamefont {Mirza}, \citenamefont {Cooper},\ and\ \citenamefont
  {Tallon}}]{Loram1998}%
  \BibitemOpen
  \bibfield  {author} {\bibinfo {author} {\bibfnamefont {J.~W.}\ \bibnamefont
  {Loram}}, \bibinfo {author} {\bibfnamefont {K.~A.}\ \bibnamefont {Mirza}},
  \bibinfo {author} {\bibfnamefont {J.~R.}\ \bibnamefont {Cooper}},\ and\
  \bibinfo {author} {\bibfnamefont {J.~L.}\ \bibnamefont {Tallon}},\ }\bibfield
   {title} {\bibinfo {title} {{SPECIFIC HEAT EVIDENCE ON THE NORMAL STATE
  PSEUDOGAP}},\ }\href {https://doi.org/10.1016/S0022-3697(98)00180-2}
  {\bibfield  {journal} {\bibinfo  {journal} {J. Phys. Chem. Solids}\ }\textbf
  {\bibinfo {volume} {59}},\ \bibinfo {pages} {2091} (\bibinfo {year}
  {1998})}\BibitemShut {NoStop}%
\bibitem [{\citenamefont {Michon}\ \emph {et~al.}(2019)\citenamefont {Michon},
  \citenamefont {Girod}, \citenamefont {Badoux}, \citenamefont {Ka{\v c}mar{\v
  c}{\'\i}k}, \citenamefont {Ma}, \citenamefont {Dragomir}, \citenamefont
  {Dabkowska}, \citenamefont {Gaulin}, \citenamefont {Zhou}, \citenamefont
  {Pyon}, \citenamefont {Takayama}, \citenamefont {Takagi}, \citenamefont
  {Verret}, \citenamefont {Doiron-Leyraud}, \citenamefont {Marcenat},
  \citenamefont {Taillefer},\ and\ \citenamefont {Klein}}]{Michon2019}%
  \BibitemOpen
  \bibfield  {author} {\bibinfo {author} {\bibfnamefont {B.}~\bibnamefont
  {Michon}}, \bibinfo {author} {\bibfnamefont {C.}~\bibnamefont {Girod}},
  \bibinfo {author} {\bibfnamefont {S.}~\bibnamefont {Badoux}}, \bibinfo
  {author} {\bibfnamefont {J.}~\bibnamefont {Ka{\v c}mar{\v c}{\'\i}k}},
  \bibinfo {author} {\bibfnamefont {Q.}~\bibnamefont {Ma}}, \bibinfo {author}
  {\bibfnamefont {M.}~\bibnamefont {Dragomir}}, \bibinfo {author}
  {\bibfnamefont {H.~A.}\ \bibnamefont {Dabkowska}}, \bibinfo {author}
  {\bibfnamefont {B.~D.}\ \bibnamefont {Gaulin}}, \bibinfo {author}
  {\bibfnamefont {J.-S.}\ \bibnamefont {Zhou}}, \bibinfo {author}
  {\bibfnamefont {S.}~\bibnamefont {Pyon}}, \bibinfo {author} {\bibfnamefont
  {T.}~\bibnamefont {Takayama}}, \bibinfo {author} {\bibfnamefont
  {H.}~\bibnamefont {Takagi}}, \bibinfo {author} {\bibfnamefont
  {S.}~\bibnamefont {Verret}}, \bibinfo {author} {\bibfnamefont
  {N.}~\bibnamefont {Doiron-Leyraud}}, \bibinfo {author} {\bibfnamefont
  {C.}~\bibnamefont {Marcenat}}, \bibinfo {author} {\bibfnamefont
  {L.}~\bibnamefont {Taillefer}},\ and\ \bibinfo {author} {\bibfnamefont
  {T.}~\bibnamefont {Klein}},\ }\bibfield  {title} {\bibinfo {title}
  {Thermodynamic signatures of quantum criticality in cuprate
  superconductors},\ }\href {https://doi.org/10.1038/s41586-019-0932-x}
  {\bibfield  {journal} {\bibinfo  {journal} {Nature}\ }\textbf {\bibinfo
  {volume} {567}},\ \bibinfo {pages} {218} (\bibinfo {year}
  {2019})}\BibitemShut {NoStop}%
\bibitem [{\citenamefont {Bari{\v s}i{\'c}}\ \emph {et~al.}(2019)\citenamefont
  {Bari{\v s}i{\'c}}, \citenamefont {Chan}, \citenamefont {Veit}, \citenamefont
  {Dorow}, \citenamefont {Ge}, \citenamefont {Li}, \citenamefont {Tabis},
  \citenamefont {Tang}, \citenamefont {Yu}, \citenamefont {Zhao},\ and\
  \citenamefont {Greven}}]{Barisic2019}%
  \BibitemOpen
  \bibfield  {author} {\bibinfo {author} {\bibfnamefont {N.}~\bibnamefont
  {Bari{\v s}i{\'c}}}, \bibinfo {author} {\bibfnamefont {M.~K.}\ \bibnamefont
  {Chan}}, \bibinfo {author} {\bibfnamefont {M.~J.}\ \bibnamefont {Veit}},
  \bibinfo {author} {\bibfnamefont {C.~J.}\ \bibnamefont {Dorow}}, \bibinfo
  {author} {\bibfnamefont {Y.}~\bibnamefont {Ge}}, \bibinfo {author}
  {\bibfnamefont {Y.}~\bibnamefont {Li}}, \bibinfo {author} {\bibfnamefont
  {W.}~\bibnamefont {Tabis}}, \bibinfo {author} {\bibfnamefont
  {Y.}~\bibnamefont {Tang}}, \bibinfo {author} {\bibfnamefont {G.}~\bibnamefont
  {Yu}}, \bibinfo {author} {\bibfnamefont {X.}~\bibnamefont {Zhao}},\ and\
  \bibinfo {author} {\bibfnamefont {M.}~\bibnamefont {Greven}},\ }\bibfield
  {title} {\bibinfo {title} {{Evidence for a universal Fermi-liquid scattering
  rate throughout the phase diagram of the copper-oxide superconductors}},\
  }\href {https://doi.org/10.1088/1367-2630/ab4d0f} {\bibfield  {journal}
  {\bibinfo  {journal} {New J. Phys.}\ }\textbf {\bibinfo {volume} {21}},\
  \bibinfo {pages} {113007} (\bibinfo {year} {2019})}\BibitemShut {NoStop}%
\bibitem [{\citenamefont {Bandur}\ \emph {et~al.}(2025)\citenamefont {Bandur},
  \citenamefont {Lee}, \citenamefont {Tsankov}, \citenamefont {Erb},\ and\
  \citenamefont {Haase}}]{Bandur2025}%
  \BibitemOpen
  \bibfield  {author} {\bibinfo {author} {\bibfnamefont {D.}~\bibnamefont
  {Bandur}}, \bibinfo {author} {\bibfnamefont {A.}~\bibnamefont {Lee}},
  \bibinfo {author} {\bibfnamefont {S.}~\bibnamefont {Tsankov}}, \bibinfo
  {author} {\bibfnamefont {A.}~\bibnamefont {Erb}},\ and\ \bibinfo {author}
  {\bibfnamefont {J.}~\bibnamefont {Haase}},\ }\bibfield  {title} {\bibinfo
  {title} {Stripe-like correlations in the cuprates from oxygen {NMR}},\ }\href
  {https://doi.org/10.1016/j.physc.2025.1354722} {\bibfield  {journal}
  {\bibinfo  {journal} {Physica C: Superconductivity and its Applications}\
  }\textbf {\bibinfo {volume} {633}},\ \bibinfo {pages} {1354722} (\bibinfo
  {year} {2025})}\BibitemShut {NoStop}%
\bibitem [{\citenamefont {Li}\ \emph {et~al.}(2023)\citenamefont {Li},
  \citenamefont {Huang}, \citenamefont {Ren}, \citenamefont {Weschke},
  \citenamefont {Ju}, \citenamefont {Zou}, \citenamefont {Zhang}, \citenamefont
  {Qiu}, \citenamefont {Liu}, \citenamefont {Ding}, \citenamefont {Singh},
  \citenamefont {Prokhnenko}, \citenamefont {Huang}, \citenamefont {Esterlis},
  \citenamefont {Wang}, \citenamefont {Xie},\ and\ \citenamefont
  {Peng}}]{Li2023}%
  \BibitemOpen
  \bibfield  {author} {\bibinfo {author} {\bibfnamefont {Q.}~\bibnamefont
  {Li}}, \bibinfo {author} {\bibfnamefont {H.-Y.}\ \bibnamefont {Huang}},
  \bibinfo {author} {\bibfnamefont {T.}~\bibnamefont {Ren}}, \bibinfo {author}
  {\bibfnamefont {E.}~\bibnamefont {Weschke}}, \bibinfo {author} {\bibfnamefont
  {L.}~\bibnamefont {Ju}}, \bibinfo {author} {\bibfnamefont {C.}~\bibnamefont
  {Zou}}, \bibinfo {author} {\bibfnamefont {S.}~\bibnamefont {Zhang}}, \bibinfo
  {author} {\bibfnamefont {Q.}~\bibnamefont {Qiu}}, \bibinfo {author}
  {\bibfnamefont {J.}~\bibnamefont {Liu}}, \bibinfo {author} {\bibfnamefont
  {S.}~\bibnamefont {Ding}}, \bibinfo {author} {\bibfnamefont {A.}~\bibnamefont
  {Singh}}, \bibinfo {author} {\bibfnamefont {O.}~\bibnamefont {Prokhnenko}},
  \bibinfo {author} {\bibfnamefont {D.-J.}\ \bibnamefont {Huang}}, \bibinfo
  {author} {\bibfnamefont {I.}~\bibnamefont {Esterlis}}, \bibinfo {author}
  {\bibfnamefont {Y.}~\bibnamefont {Wang}}, \bibinfo {author} {\bibfnamefont
  {Y.}~\bibnamefont {Xie}},\ and\ \bibinfo {author} {\bibfnamefont
  {Y.}~\bibnamefont {Peng}},\ }\bibfield  {title} {\bibinfo {title}
  {{Prevailing Charge Order in Overdoped La$_{2-x}$Sr$_x$CuO$_4$ beyond the
  Superconducting Dome}},\ }\href
  {https://doi.org/10.1103/PhysRevLett.131.116002} {\bibfield  {journal}
  {\bibinfo  {journal} {Phys. Rev. Lett.}\ }\textbf {\bibinfo {volume} {131}},\
  \bibinfo {pages} {116002} (\bibinfo {year} {2023})}\BibitemShut {NoStop}%
\bibitem [{\citenamefont {Tranquada}\ \emph {et~al.}(1995)\citenamefont
  {Tranquada}, \citenamefont {Sternlieb}, \citenamefont {Axe}, \citenamefont
  {Nakamura},\ and\ \citenamefont {Uchida}}]{Tranquada1995}%
  \BibitemOpen
  \bibfield  {author} {\bibinfo {author} {\bibfnamefont {J.~M.}\ \bibnamefont
  {Tranquada}}, \bibinfo {author} {\bibfnamefont {B.~J.}\ \bibnamefont
  {Sternlieb}}, \bibinfo {author} {\bibfnamefont {J.~D.}\ \bibnamefont {Axe}},
  \bibinfo {author} {\bibfnamefont {Y.}~\bibnamefont {Nakamura}},\ and\
  \bibinfo {author} {\bibfnamefont {S.}~\bibnamefont {Uchida}},\ }\bibfield
  {title} {\bibinfo {title} {{Evidence for stripe correlations of spins and
  holes in copper oxide superconductors}},\ }\href
  {https://doi.org/10.1038/375561a0} {\bibfield  {journal} {\bibinfo  {journal}
  {Nature}\ }\textbf {\bibinfo {volume} {375}},\ \bibinfo {pages} {561}
  (\bibinfo {year} {1995})}\BibitemShut {NoStop}%
\bibitem [{\citenamefont {Keller}\ \emph {et~al.}(2008)\citenamefont {Keller},
  \citenamefont {Bussmann-Holder},\ and\ \citenamefont
  {M{\"u}ller}}]{Keller2008}%
  \BibitemOpen
  \bibfield  {author} {\bibinfo {author} {\bibfnamefont {H.}~\bibnamefont
  {Keller}}, \bibinfo {author} {\bibfnamefont {A.}~\bibnamefont
  {Bussmann-Holder}},\ and\ \bibinfo {author} {\bibfnamefont {K.~A.}\
  \bibnamefont {M{\"u}ller}},\ }\bibfield  {title} {\bibinfo {title}
  {{Jahn-Teller physics and high-$T_c$ superconductivity}},\ }\href
  {https://doi.org/10.1016/S1369-7021(08)70178-0} {\bibfield  {journal}
  {\bibinfo  {journal} {Materials Today}\ }\textbf {\bibinfo {volume} {11}},\
  \bibinfo {pages} {38} (\bibinfo {year} {2008})}\BibitemShut {NoStop}%
\bibitem [{\citenamefont {Bianconi}\ \emph {et~al.}(1996)\citenamefont
  {Bianconi}, \citenamefont {Saini}, \citenamefont {Lanzara}, \citenamefont
  {Missori}, \citenamefont {Rossetti}, \citenamefont {Oyanagi}, \citenamefont
  {Yamaguchi}, \citenamefont {Oka},\ and\ \citenamefont {Ito}}]{Bianconi1996}%
  \BibitemOpen
  \bibfield  {author} {\bibinfo {author} {\bibfnamefont {A.}~\bibnamefont
  {Bianconi}}, \bibinfo {author} {\bibfnamefont {N.~L.}\ \bibnamefont {Saini}},
  \bibinfo {author} {\bibfnamefont {A.}~\bibnamefont {Lanzara}}, \bibinfo
  {author} {\bibfnamefont {M.}~\bibnamefont {Missori}}, \bibinfo {author}
  {\bibfnamefont {T.}~\bibnamefont {Rossetti}}, \bibinfo {author}
  {\bibfnamefont {H.}~\bibnamefont {Oyanagi}}, \bibinfo {author} {\bibfnamefont
  {H.}~\bibnamefont {Yamaguchi}}, \bibinfo {author} {\bibfnamefont
  {K.}~\bibnamefont {Oka}},\ and\ \bibinfo {author} {\bibfnamefont
  {T.}~\bibnamefont {Ito}},\ }\bibfield  {title} {\bibinfo {title}
  {{Determination of the Local Lattice Distortions in the Cu${\mathrm{O}}_{2}$
  Plane of L${\mathrm{a}}_{1.85}$S${\mathrm{r}}_{0.15}$Cu${\mathrm{O}}_{4}$}},\
  }\href {https://doi.org/10.1103/PhysRevLett.76.3412} {\bibfield  {journal}
  {\bibinfo  {journal} {Phys. Rev. Lett.}\ }\textbf {\bibinfo {volume} {76}},\
  \bibinfo {pages} {3412} (\bibinfo {year} {1996})}\BibitemShut {NoStop}%
\bibitem [{\citenamefont {Bussmann-Holder}\ and\ \citenamefont
  {Keller}(2024)}]{Bussmann2024}%
  \BibitemOpen
  \bibfield  {author} {\bibinfo {author} {\bibfnamefont {A.}~\bibnamefont
  {Bussmann-Holder}}\ and\ \bibinfo {author} {\bibfnamefont {H.}~\bibnamefont
  {Keller}},\ }\bibfield  {title} {\bibinfo {title} {{Multiband
  Superconductivity, Polarons, Jahn-Teller Polarons, Heterogeneity, and
  High-Temperature Superconductivity}},\ }\href
  {https://doi.org/10.3390/condmat9040056} {\bibfield  {journal} {\bibinfo
  {journal} {Condensed Matter}\ }\textbf {\bibinfo {volume} {9}},\ \bibinfo
  {pages} {56} (\bibinfo {year} {2024})}\BibitemShut {NoStop}%
\bibitem [{\citenamefont {Gor'kov}(2001)}]{Gorkov2001}%
  \BibitemOpen
  \bibfield  {author} {\bibinfo {author} {\bibfnamefont {L.~P.}\ \bibnamefont
  {Gor'kov}},\ }\bibfield  {title} {\bibinfo {title} {Inherent {Inhomogeneity}
  in {Two}-{Component} {Model} for {Cuprates}},\ }\href
  {https://doi.org/10.1023/A:1011183504208} {\bibfield  {journal} {\bibinfo
  {journal} {Journal of Superconductivity}\ }\textbf {\bibinfo {volume} {14}},\
  \bibinfo {pages} {365} (\bibinfo {year} {2001})}\BibitemShut {NoStop}%
\bibitem [{\citenamefont {Suhl}\ \emph {et~al.}(1959)\citenamefont {Suhl},
  \citenamefont {Matthias},\ and\ \citenamefont {Walker}}]{Suhl1959}%
  \BibitemOpen
  \bibfield  {author} {\bibinfo {author} {\bibfnamefont {H.}~\bibnamefont
  {Suhl}}, \bibinfo {author} {\bibfnamefont {B.~T.}\ \bibnamefont {Matthias}},\
  and\ \bibinfo {author} {\bibfnamefont {L.~R.}\ \bibnamefont {Walker}},\
  }\bibfield  {title} {\bibinfo {title} {{Bardeen-Cooper-Schrieffer Theory of
  Superconductivity in the Case of Overlapping Bands}},\ }\href
  {https://doi.org/10.1103/PhysRevLett.3.552} {\bibfield  {journal} {\bibinfo
  {journal} {Phys. Rev. Lett.}\ }\textbf {\bibinfo {volume} {3}},\ \bibinfo
  {pages} {552} (\bibinfo {year} {1959})}\BibitemShut {NoStop}%
\bibitem [{\citenamefont {Matt}\ \emph {et~al.}(2018)\citenamefont {Matt},
  \citenamefont {Sutter}, \citenamefont {Cook}, \citenamefont {Sassa},
  \citenamefont {M{\aa}nsson}, \citenamefont {Tjernberg}, \citenamefont {Das},
  \citenamefont {Horio}, \citenamefont {Destraz}, \citenamefont {Fatuzzo},
  \citenamefont {Hauser}, \citenamefont {Shi}, \citenamefont {Kobayashi},
  \citenamefont {Strocov}, \citenamefont {Schmitt}, \citenamefont {Dudin},
  \citenamefont {Hoesch}, \citenamefont {Pyon}, \citenamefont {Takayama},
  \citenamefont {Takagi}, \citenamefont {Lipscombe}, \citenamefont {Hayden},
  \citenamefont {Kurosawa}, \citenamefont {Momono}, \citenamefont {Oda},
  \citenamefont {Neupert},\ and\ \citenamefont {Chang}}]{Matt2018}%
  \BibitemOpen
  \bibfield  {author} {\bibinfo {author} {\bibfnamefont {C.~E.}\ \bibnamefont
  {Matt}}, \bibinfo {author} {\bibfnamefont {D.}~\bibnamefont {Sutter}},
  \bibinfo {author} {\bibfnamefont {A.~M.}\ \bibnamefont {Cook}}, \bibinfo
  {author} {\bibfnamefont {Y.}~\bibnamefont {Sassa}}, \bibinfo {author}
  {\bibfnamefont {M.}~\bibnamefont {M{\aa}nsson}}, \bibinfo {author}
  {\bibfnamefont {O.}~\bibnamefont {Tjernberg}}, \bibinfo {author}
  {\bibfnamefont {L.}~\bibnamefont {Das}}, \bibinfo {author} {\bibfnamefont
  {M.}~\bibnamefont {Horio}}, \bibinfo {author} {\bibfnamefont
  {D.}~\bibnamefont {Destraz}}, \bibinfo {author} {\bibfnamefont {C.~G.}\
  \bibnamefont {Fatuzzo}}, \bibinfo {author} {\bibfnamefont {K.}~\bibnamefont
  {Hauser}}, \bibinfo {author} {\bibfnamefont {M.}~\bibnamefont {Shi}},
  \bibinfo {author} {\bibfnamefont {M.}~\bibnamefont {Kobayashi}}, \bibinfo
  {author} {\bibfnamefont {V.~N.}\ \bibnamefont {Strocov}}, \bibinfo {author}
  {\bibfnamefont {T.}~\bibnamefont {Schmitt}}, \bibinfo {author} {\bibfnamefont
  {P.}~\bibnamefont {Dudin}}, \bibinfo {author} {\bibfnamefont
  {M.}~\bibnamefont {Hoesch}}, \bibinfo {author} {\bibfnamefont
  {S.}~\bibnamefont {Pyon}}, \bibinfo {author} {\bibfnamefont {T.}~\bibnamefont
  {Takayama}}, \bibinfo {author} {\bibfnamefont {H.}~\bibnamefont {Takagi}},
  \bibinfo {author} {\bibfnamefont {O.~J.}\ \bibnamefont {Lipscombe}}, \bibinfo
  {author} {\bibfnamefont {S.~M.}\ \bibnamefont {Hayden}}, \bibinfo {author}
  {\bibfnamefont {T.}~\bibnamefont {Kurosawa}}, \bibinfo {author}
  {\bibfnamefont {N.}~\bibnamefont {Momono}}, \bibinfo {author} {\bibfnamefont
  {M.}~\bibnamefont {Oda}}, \bibinfo {author} {\bibfnamefont {T.}~\bibnamefont
  {Neupert}},\ and\ \bibinfo {author} {\bibfnamefont {J.}~\bibnamefont
  {Chang}},\ }\bibfield  {title} {\bibinfo {title} {Direct observation of
  orbital hybridisation in a cuprate superconductor},\ }\href
  {https://doi.org/10.1038/s41467-018-03266-0} {\bibfield  {journal} {\bibinfo
  {journal} {Nature Commun.}\ }\textbf {\bibinfo {volume} {9}},\ \bibinfo
  {pages} {972} (\bibinfo {year} {2018})}\BibitemShut {NoStop}%
\bibitem [{\citenamefont {Weber}\ \emph {et~al.}(2014)\citenamefont {Weber},
  \citenamefont {Giamarchi},\ and\ \citenamefont {Varma}}]{Weber2014}%
  \BibitemOpen
  \bibfield  {author} {\bibinfo {author} {\bibfnamefont {C.}~\bibnamefont
  {Weber}}, \bibinfo {author} {\bibfnamefont {T.}~\bibnamefont {Giamarchi}},\
  and\ \bibinfo {author} {\bibfnamefont {C.~M.}\ \bibnamefont {Varma}},\
  }\bibfield  {title} {\bibinfo {title} {Phase diagram of a three-orbital model
  for high-${T}_{c}$ cuprate superconductors},\ }\href
  {https://doi.org/10.1103/PhysRevLett.112.117001} {\bibfield  {journal}
  {\bibinfo  {journal} {Phys. Rev. Lett.}\ }\textbf {\bibinfo {volume} {112}},\
  \bibinfo {pages} {117001} (\bibinfo {year} {2014})}\BibitemShut {NoStop}%
\bibitem [{\citenamefont {Bourges}\ \emph {et~al.}(2022)\citenamefont
  {Bourges}, \citenamefont {Bounoua},\ and\ \citenamefont
  {Sidis}}]{Bourges2022}%
  \BibitemOpen
  \bibfield  {author} {\bibinfo {author} {\bibfnamefont {P.}~\bibnamefont
  {Bourges}}, \bibinfo {author} {\bibfnamefont {D.}~\bibnamefont {Bounoua}},\
  and\ \bibinfo {author} {\bibfnamefont {Y.}~\bibnamefont {Sidis}},\ }\bibfield
   {title} {\bibinfo {title} {Loop currents in quantum matter},\ }\href
  {https://doi.org/10.5802/crphys.84} {\bibfield  {journal} {\bibinfo
  {journal} {Comptes Rendus Physique}\ }\textbf {\bibinfo {volume} {22}},\
  \bibinfo {pages} {7} (\bibinfo {year} {2022})}\BibitemShut {NoStop}%
\end{thebibliography}%

\printindex

\end{document}